\documentclass[a4paper,fleqn,usenatbib,useAMS]{mnras}

\usepackage{newtxtext,newtxmath}

\usepackage[T1]{fontenc}
\usepackage{ae,aecompl}

\usepackage[dvipdfmx]{graphicx}	
\usepackage{amsmath}	
\usepackage{amssymb}	
\usepackage{bm}
\usepackage{color}

\title[Line transfer at SNR shocks: 2s-state hydrogen atoms]{Radiative transfer of hydrogen lines from supernova remnant shock waves:
contributions of 2s-state hydrogen atoms}

\author[J. Shimoda \& J. M. Laming]{
Jiro Shimoda$^{1,2}$\thanks{E-mail: j-shimoda@astr.tohoku.ac.jp (JS)}
and J. Martin Laming $^3$\thanks{E-mail: laming@nrl.navy.mil (JML)}
\\
$^{1}$Frontier Research Institute for Interdisciplinary Sciences, Tohoku University, Sendai 980-8578, Japan\\
$^{2}$Astronomical Institute, Tohoku University, Sendai 980-8578, Japan\\
$^{3}$Space Science Division Code 7684, Naval Research Laboratory, Washington DC 20375, USA\\
}

\date{Accepted XXX. Received YYY; in original form ZZZ}

\pubyear{2018}

\begin{document}
\label{firstpage}
\pagerange{\pageref{firstpage}--\pageref{lastpage}}
\maketitle

\begin{abstract}
Radiative transfer in hydrogen lines in supernova remnant (SNR) shock waves
is studied taking into account the population of the hydrogen atom
2s-state. Measurements of Balmer line emission, especially of H~$\alpha$,
are often relied upon to derive physical conditions in the SNR shock. On
the other hand, Lyman series photons, especially Ly~$\beta$, are mostly
absorbed by upstream hydrogen atoms. As a result, atoms are excited to the
3p state, and then emit H~$\alpha$ by the spontaneous transition from 3p to
2s. Thus, the nature of H~$\alpha$ depends on how many Ly~$\beta$ photons
are converted to H~$\alpha$ photons. Moreover, the Balmer lines can be
scattered by the 2s-state hydrogen atoms, which are excited not only by
collisional excitation but also by the Lyman-Balmer conversion. It is shown
for example that the H~$\alpha$ photons are scattered if the shock
propagates into an H~$_{\rm I}$ cloud with a density of $\sim30~{\rm
cm^{-3}}$ and a size of $\sim 1$~pc. We find that the line profile of
H~$\alpha$ becomes asymmetric resulting from the difference between line
centre frequencies among the transitions from 3s to 2p, from 3p to 2s and
from 3d to 2p. We also find that the broad-to-narrow ratio of H~$\alpha$,
which is often used to estimate the ion-electron temperature equilibrium,
varies at most $\simeq 10$ per cent depending on the
ionization degree of the upstream medium because of incomplete conversion
of Lyman lines to Balmer lines.
\end{abstract}

\begin{keywords}
acceleration of particles
-- atomic processes
-- radiative transfer
-- shock waves
-- cosmic rays
-- ISM: supernova remnants.
\end{keywords}



\section{Introduction}
\label{sec:introduction} Balmer line emissions from supernova remnant (SNR)
shock waves are relied upon as a probe of the physics of collisionless
shocks. The shock transition occurs on a length scale much shorter than that
associated with a particle mean free path to Coulomb scattering, so thermal
equilibrium is much less strongly enforced. Balmer lines can be used to
diagnose the resulting departures from equilibrium, such as the effects on
ion-electron temperature ratio, the nature of the shock precursor and
the acceleration of non-thermal particles~\citep[see][and
Section~\ref{sec:BDS} for reviews]{raymond91,heng10}. Such a shock is often
called a Balmer dominated shock (BDS).

The physics of particle acceleration may be the most important issue because
it is crucial to the origin of cosmic-rays. Moreover, accelerated particles
in collisionless shocks are often considered to be responsible for the
radiation from high-energy astrophysical sources from the radio to the TeV
band. An important concern is to specify the density of accelerated particles
in SNR shocks, a necessary step towards confirming SNR shocks as the main
sites producing Galactic cosmic-rays. In addition, it allows us to quantify
the back reaction of accelerated particles on the background shock structure.
Note that in the standard model of particle acceleration (i.e. diffusive
shock acceleration), the distribution function of accelerated particles
depends on the shock structure~\citep[e.g.][and references
therein]{berezhko99}. If the amount of accelerated particles is significant,
in other words, if the kinetic energy of the shock consumed due to the
particle acceleration is a large fraction of the shock energy, the downstream
temperature becomes considerably lower than the case of an adiabatic shock,
that is, there is some missing thermal energy. This energy loss from the
shock has been widely
investigated~\citep[e.g.][]{hughes00,tatischeff07,helder09,morlino13b,morlino13,morlino14,shimoda15,hovey18}.
In the latest development, \citet{shimoda18} showed that the linear
polarization degree of H~$\alpha$ observed perpendicularly to the shock
velocity vector depends on the energy-loss from the shock. Such polarized
H~$\alpha$ was originally predicted by~\citet{laming90} to estimate the
ion-electron temperature equilibrium for the adiabatic shock and was recently
discovered in SN~1006 by~\citet{sparks15}.
\par
As well as being directly excited from the ground state, Balmer lines
(intensity, line profile, polarization and so on) are affected by the
conversion of Lyman lines to Balmer lines. For example, the absorption of
Ly~$\beta$ by a hydrogen atom results in radiative excitation from 1s to 3p,
and the excited atom can emit H~$\alpha$ by the spontaneous transition from
3p to 2s. Simultaneously, the conversion yields the 2s-state hydrogen atom,
which creates the two-photon continuum by the spontaneous transition from 2s
to 1s. Thus, the Ly~$\beta$ to H~$\alpha$ conversion impacts the total
intensity, line profile and net polarization of H~$\alpha$. Moreover, an
adequate density of 2s-sate atoms can further scatter H~$\alpha$ photons.
Although such fundamental physics is well known, it has not been well studied
in SNR shocks. In fact, it is usually assumed that the Ly~$\beta$ photons are
at the limits of either completely optically thick or optically thin at SNR
shocks, that is, they are completely converted to H~$\alpha$ photons or not
at all~\citep[e.g.][]{heng07,vanAdelsberg08,morlino12,morlino13,shimoda18}.
Contrary to this, \citet{ghavamian01} studied the conversion of Ly~$\beta$
and Ly~$\gamma$ to H~$\alpha$ and H~$\beta$ by Monte Carlo simulations and
claimed that intermediate conversion occurs. However, they and previous
studies did not consider the population of 2s-state hydrogen atoms. In this
paper, we provide a formulation of the radiative line transfer with the rate
equation of atomic population and study the nature of Balmer line emissions
from SNR shocks. In this paper we do not consider the polarization, deferring
that instead to a later work. Note that as a first step, our model makes
several simplifications in the treatment of the SNR shock, handling the
hydrogen atoms as fluids and supposing no particles leaking back upstream
(e.g. cosmic-rays). Our calculation of radiative transfer is based on
so-called the ray-tracing method and uses updated atomic data from the
literature~\citep[e.g.][]{heng08,tseli12}. Moreover, we consider only
hydrogen line emissions and ignore bremsstrahlung radiation, thermal
emissions from the SNR ejecta, and external radiation sources. Thus, our
model possibly predicts somewhat smaller population of 2s-state hydrogen
atoms than would be the case in a realistic SNR shock.
\par
This paper is organized as follows. In Section~\ref{sec:BDS}, we briefly
review the BDSs as a probe of the collisionless shock physics and give a
simple estimate of the occupation number of the 2s-state hydrogen atoms. In
Section~\ref{sec:formulation}, we formulate the radiative line transfer
problem for the SNR shock. In Section~\ref{sec:population}, we present the
results for atomic populations. In Section~\ref{sec:observation}, we consider
how the hydrogen lines are observed based on the results of the atomic
population computations. Finally, we summarize our
results.

\section{Diagnostics of Balmer Dominated Shocks}
\label{sec:BDS} In this section, we will briefly review the diagnostics of
BDSs and estimate the previously neglected population of 2s-state hydrogen
atoms.
\par
The basic theory of the Balmer line emission from SNR shocks was described by
\citet{chevalier80}. They pointed out that hydrogen atoms in the interstellar
medium (ISM) do not suffer shock-heating across the shock front because the
SNR shock is `collisionless'. The shock transition is formed by the
interaction between charged particles and plasma waves rather than by Coulomb
collisions. Then, the hydrogen atoms entering the downstream region collide
with the shock-heated, charged particles. The collisions result in several
atomic reactions such as ionization, direct excitation and charge-transfer
between the hydrogen atoms and the shock-heated protons. Atoms which have not
experienced any charge-transfer reactions emit a `narrow' line with width of
$\sim10~{\rm km~s^{-1}}$, while the atoms that have undergone a
charge-transfer reaction emit a `broad' line with a width characteristic of
the post-shock ion temperature; $\sim1000~{\rm km~s^{-1}}$. Thus, the profile
of hydrogen lines observed in the SNR shocks consists of at least these two
components. The width of each component corresponds to the upstream
temperature (narrow) and the downstream proton temperature (broad),
respectively. Thus, the downstream temperature can be derived from the width
of broad component. The intensity ratio of the broad component to the narrow
component is often relied on to estimate the ion-electron temperature
ratio~\citep{ghavamian13}. The Balmer decrement, which
is the intensity ratio of the narrow H~$\alpha$ to the narrow H~$\beta$,
depends on the ionization degree of the ambient gas around the SNR shock.
The H~$\alpha$ (H~$\beta$) intensity is enhanced via the
conversion from Ly~$\beta$ (Ly~$\gamma$) to H~$\alpha$ (H~$\beta$). Since the
absorption cross-section of Ly~$\beta$ is larger than that of Ly~$\gamma$,
the Ly~$\beta$ to H~$\alpha$ conversion may occur more than the Ly~$\gamma$
to H~$\beta$ conversion. Thus, the Balmer decrement reflects the optical
properties of BDSs, which depends on the ionization structure of hydrogen. 
This basic model did not consider the existence of particles leaking back
upstream (e.g. cosmic-rays and fast-neutral particles produced by the
charge-transfer reaction) and was further developed in several
papers~\citep[e.g.][]{ghavamian01,heng07,vanAdelsberg08}.
\par
The existence of particles leaking back to the upstream region is implied by
observations of the full width at half maximum (FWHM) of the narrow component
of H~$\alpha$ (30--50~km~s$^{-1}$) and H~$\alpha$ emission from the upstream
region; i.e. the shock-precursor emission
\citep[e.g.][]{smith94,ghavamian00,sollerman03,lee07,lee10,medina14,katsuda16,knezevic17}.
Note that the FWHM of 30--50~km~s$^{-1}$ implies an upstream temperature of
2.5--5.6~eV, which would be too high for neutral hydrogen atoms to exist if
it was the equilibrium temperature of the ISM. Therefore, non-thermal
pre-heating by the leaking particles in the upstream region is expected.
Moreover, \citet{raymond10} showed that the H~$\alpha$ line profile observed
in {\it Tycho's} SNR can be fitted by three Gaussian functions. This implies
that there is an intermediate component with temperature between the narrow
and broad components (i.e. a non-thermal velocity distribution of hydrogen
atoms undergoing the charge-transfer reaction) or the
velocity distribution of downstream protons undergoing charge-transfer
reaction deviates from the Maxwellian distribution due to a non-thermal
wing~\citep{raymond10,raymond17}. Semi-analytical
models~\citep{morlino12,morlino13} and hybrid simulations~\citep{ohira16}
suggest that the cosmic-rays and/or the fast-neutral particles emerging from
the charge-transfer reaction lead to such shock-precursor emission
accompanied by the intermediate component and the anomalous width of narrow
component. Note that the shock-precursor emission due to the absorption of
Lyman photons may be inevitable. In this paper, we refer to such precursor
emission as `shock-precursor-like emission' or `photo-precursor emission' to
distinguish it from the cosmic-ray/fast-neutral precursor.
\par
Here we give a simple estimate for the population of 2s-state hydrogen atoms
by considering a 3-level system (1s, 2s and 3p). We set the rate equation for
the 2s population as%
\begin{eqnarray}
n_{\rm H,1s}C_{\rm 1s,2s}+n_{\rm H,3p}A_{\rm 3p,2s}-n_{\rm H,2s}A_{\rm 2s,1s}=0,
\end{eqnarray}
%
where $n_{{\rm H},j}$ is the number density of hydrogen atoms in the state
$j$, $C_{j,k}$ and $A_{j,k}$ are the collisional excitation rate and
spontaneous decay rate for the transition from $j$ to $k$, respectively. For
the bound states, we use the notation $j=n_j l_j$, where $n_j$ is the
principal quantum number of the state $j$. Similarly,
$l_j=0,1,2,3,...,n_j-1$~(equivalently: s,~p,~d,~f,...) is the orbital
angular-momentum quantum number of the state $j$. Here we
suppose that the depopulation term of 2s-state atoms is dominated by the
spontaneous transition at the rate of $A_{\rm 2s,1s}\simeq8.2~{\rm s^{-1}}$.
In reality, the collisional transition from 2s to 2p can be a subdominant
process for depopulation. For a collision at a velocity $\sim10^8~{\rm
cm~s^{-1}}$, which is a typical velocity scale for young SNR shocks, the
cross-section is $\sim10^{-13}~{\rm cm^2}$, giving a reaction rate
$\sim10^{-5}~{\rm cm^3~s^{-1}}$~\citep[e.g.][]{janev87,sahal96}. Thus, if the
density is $\sim10^6~{\rm cm^{-3}}$, the collisional depopulation becomes
important. Note that we assume no strong radiation field inducing the
radiative transition from 2s to any other state.\footnote{
The cross-section of 2s-2p collisional excitation has the
maximum value $\sim10^{-10}~{\rm cm^2}$ around the relative velocity
$\sim10^6\mathchar`-10^7~{\rm cm~s^{-1}}$ for both proton and electron
impacts~\citep{janev87}, leading to a rate of $10^{-4} - 10^{-3}$
cm$^3$s$^{-1}$ and critical densities of $10^4 - 10^5$ cm$^{-3}$ at slower
shocks.} The occupation number of 3p, $n_{\rm H,3p}$, depends on the
absorption of Ly~$\beta$. Here we assume an isotropic radiation field for
Ly~$\beta$. Then, we obtain the rate equation for 3p as
%
\begin{eqnarray}
n_{\rm H,1s}
\left(C_{\rm 1s,3p}+\int_0^{\infty} \frac{4\upi \sigma_\nu^{\rm 1s,3p}}{h\nu}I_\nu {\rm d}\nu \right)
-n_{\rm H,3p}
\left(A_{\rm 3p,1s}+A_{\rm 3p,2s}\right)=0,
\end{eqnarray}
%
where $h$, $\nu$, $\sigma_\nu^{\rm 1s,3p}$ and $I_\nu$ are the Planck
constant, frequency, absorption cross-section for the transition from 1s to
3p and the specific intensity, respectively. The intensity is set to be
%
\begin{eqnarray}
I_\nu=S_\nu(1-{\rm e}^{-\tau_\nu})
=\frac{ \frac{h\nu}{4\upi} A_{\rm 3p,1s} n_{\rm H,3p} }{\sigma' n_{\rm H,1s}}
(1-{\rm e}^{-\tau_\nu}),
\end{eqnarray}
%
where $S_\nu$ and $\tau_\nu$ are the source function and optical depth,
respectively. $\sigma'$ is a combination of physical constants relevant to
the radiative absorption cross-section. Thus, we derive the occupation number
of 2s as
%
\begin{eqnarray}
n_{\rm H,2s}=\frac{C_{\rm 1s,2s}}{A_{\rm 2s,1s}}
\left[
1+\frac{A_{\rm 3p,2s}}{ {\rm e}^{-\tau_0}A_{\rm 3p,1s}+A_{\rm 3p,2s} }
\frac{ C_{\rm 1s,3p} }{ C_{\rm 1s,2s} }
\right]
n_{\rm H,1s},
\label{eq:2s simple}
\end{eqnarray}
%
where $\tau_0$ is the optical depth at the line centre.
Here we assume a narrow line profile function $\phi_\nu$ for which
we can approximate as $\int_0^{\infty}(1-{\rm e}^{-\tau_\nu}){\rm d}\nu\approx1-{\rm e}^{-\tau_0}$.
The terms in the brackets
[...] indicate the contribution of the combination of the absorption and
cascades. Note that roughly say, the ratios are
$A_{\rm 3p,2s}/(A_{\rm 3p,1s}+A_{\rm 3p,2s})\simeq0.118$,
$C_{\rm 1s,3p}/C_{\rm 1s,2s}\sim2\mathchar`-10$ and $C_{\rm
1s,2s}/A_{\rm 2s,1s}\sim10^{-9}n_{\rm p}$, where $n_{\rm p}$ is the proton
number density. Thus, if Ly~$\beta$ is in the optically thick limit, $n_{\rm
H,2s}$ is enhanced roughly at most ten times compared with the optically thin
case. The absorption coefficient of H~$\alpha$ at the line centre becomes
%
\begin{eqnarray}
k_0({\rm H\alpha})
&=& \sigma_0({\rm H\alpha}) n_{\rm H,2s} \\
&\sim&
10^{-23}\mathchar`-10^{-22}~{\rm cm^{-1}}
\left(\frac{T_0}{6000~{\rm K}}\right)^{-\frac{1}{2}}
\left( \frac{n_{\rm H,1s}}{ {\rm 1~cm^{-3} } }\right)
\left( \frac{n_{\rm p}}{ {\rm 1~cm^{-3} } }\right), \nonumber
\label{eq:absp simple}
\end{eqnarray}
%
where $\sigma_0({\rm H\alpha})$ is the radiative cross-section of H~$\alpha$
at the line centre for given temperature $T_0$. Thus, if the SNR shock
interacts with somewhat dense clump with a density of $\sim30~{\rm cm^{-3}}$
and a size of $\sim 1~{\rm pc}$, the H~$\alpha$ emission can be scattered.
Note that the H~$\beta$ emission can also be scattered but its absorption
coefficient is about quarter of the H~$\alpha$ coefficient. The interaction
between the shock and a dense clump is implied by the ripple of an SNR shock
with a length-scale of $\sim10$ per cent of SNR radius~\citep[e.g.][and see
the discussion of
\citealt{shimoda15}]{ishihara10,williams13,williams16,miceli14,sano17,tsubone17}.
Note that according to magnetohydrodynamic simulations
performed by~\citet{inoue09,inoue12b}, even if the shock propagates into a
simulated ISM having density contrast ranging in $\sim1\mathchar`-30~{\rm
cm^{-3}}$ as a consequence of thermal instability, the scale length of
rippling is $\sim10$ per cent of the length of sides of simulation box.
\par
The temperature $6000$~K we assumed is often taken for
the warm neutral medium of ISM~\citep[e.g.][]{ferriere01}. If we suppose the
temperatures implied by the measured H~$\alpha$ widths of
\citet{sollerman03}, we obtain the absorption coefficients around half those
assumed at $6000$~K. Note that for SNR Cygnus Loop, \citet{medina14} pointed
out that the pre-shock gas is photoionized and heated up to $\sim17000$~K by
the emissions from post-shock region.
\par
In an actual SNR shock, the optical depth of Ly~$\beta$ may be
intermediate~\citep{ghavamian01}. Moreover, the validity of this simple
estimate is still unclear because of the many complexities of BDSs,
especially the ionization structure of hydrogen. Therefore, we consider more
sophisticated formulas for the line transfer problem and solve them
numerically.

\section{Formulation of line transfer}
\label{sec:formulation}
%
\begin{figure}
\includegraphics[scale=0.40, clip]{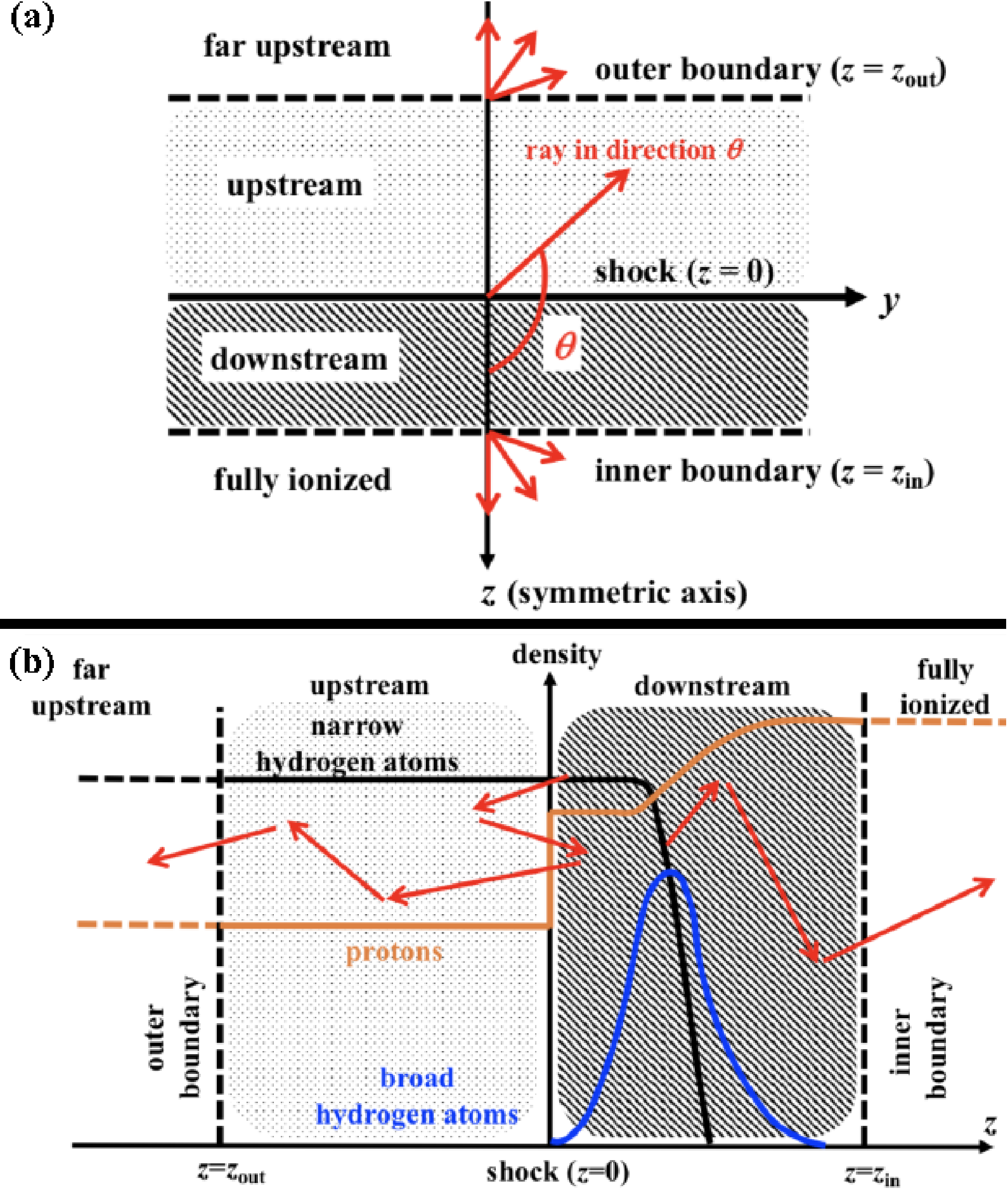}
\caption{Schematic illustrations of the SNR shock.
(a):
The shock is axially symmetric about the $z$-axis.
The $x\mathchar`-y$ plane corresponds to the shock surface.
The red arrows indicate the photon ray,
which makes an angle $\theta$ with the $z$-axis.
The upstream side is $z<0$, while the downstream side is $z>0$.
Two broken lines at $z=z_{\rm out}$ and $z=z_{\rm in}$
represent the free-escape boundaries
of photons for the upstream  and the downstream, respectively.
(b):
The curves with colors black, blue and orange indicate
the number density of narrow hydrogen atoms, broad hydrogen atoms
and protons, respectively.
Here we assume that there are no particles leaking to the upstream region.
The red arrows represent the rays of scattered photons,
which escape from the shock by crossing the outer/inner boundary.
}
\label{fig:shock}
\end{figure}
%
The line transfer problem is reviewed in several
papers~\citep[e.g.][]{castor04}. We apply their formulation to the problem
for SNR shocks propagating into pure atomic hydrogen plasma, which consists
of hydrogen atoms (denoted 'H'), protons ('p') and electrons ('e'). The shock
is set to be stationary, axially symmetric about the $z$-axis, plane-parallel
to $x\mathchar`-y$ plane and located at $z=0$ (see, Fig.~\ref{fig:shock}a).
We set two free-escape boundaries for photons upstream ($z=z_{\rm out}$) and
downstream ($z=z_{\rm in}$) of the shock. For simplicity, we assume that
there are no particles leaking to the upstream region and that the radiation
field consists of only the H line emissions (i.e. bremsstrahlung radiation,
emission from the SNR ejecta and any other external radiation sources are
neglected). Moreover, we assume temperature equilibrium for the upstream
plasma and fix the upstream temperature at $T_{\rm 0}=6000$~K for simplicity.
\par
Firstly, we describe the ionization structure of hydrogen. Let $n_{{\rm
H},j}^{\rm N}$ be the number density of 'narrow' (i.e. cold) hydrogen atoms,
which have not experienced charge-exchange reactions, while $n_{{\rm
H},j}^{\rm B}$ is the number density of 'broad' (i.e. hot) hydrogen atoms
emerging from charge-exchange reactions. Obviously, we have the relation
$n_{{\rm H},j}=n_{{\rm H},j}^{\rm N}+n_{{\rm H},j}^{\rm B}$.
Fig.~\ref{fig:shock}b is a schematic illustration of the spatial distribution
of particles. We consider that the partially ionized plasma flows from the
far upstream region ($z<z_{\rm out}$) and presume that it is in ionization
equilibrium. Hence, we set the boundary conditions as $n_{\rm H,1s}^{\rm
N}(z<0)=n_{\rm H,1s}^{\rm N}(z_{\rm out})$, $n_{\rm H,1s}^{\rm B}(z<0)=0$ and
$n_{\rm p}(z<0)=n_{\rm p}(z_{\rm out})$, where $n_{\rm p}$ is the number
density of protons. At the shock ($z=0$), we assume the strong shock jump
conditions,
%
\begin{eqnarray}
n_{\rm p}(0)
&=& 4 n_{\rm p}(z_{\rm out}), \\
u_2
&=& \frac{V_{\rm sh}}{4}, \\
k_{\rm B}T_{\rm p}
&=& \frac{3}{16}\mu' m_{\rm p}V_{\rm sh}{}^2 \\
T_{\rm e}
&=&\beta T_{\rm p}
\end{eqnarray}
%
where $V_{\rm sh}$ is the shock velocity, $k_{\rm B}$ is the Boltzmann
constant and $m_{\rm p}$ is the proton mass. $T_{\rm p}$ and $T_{\rm e}$ are
the downstream temperatures of protons and electrons, respectively. The
effective mean molecular weight is $\mu'$, which is defined as
%
\begin{eqnarray}
\mu'=1-\left(1-\mu_\odot' \right)
\frac{\beta-\frac{m_{\rm e}} {m_{\rm p}} }{\mu_\odot'+(1-\mu_\odot')\beta-\frac{m_{\rm e}}{m_{\rm p}}},
\end{eqnarray}
%
where $\mu_\odot'=0.62$ and $m_{\rm e}$ is the electron mass~\citep[see][for
details]{shimoda18}. Note that the number density of downstream protons is
function of $z$, while $u_2$, $T_{\rm p}$ and $T_{\rm e}$ are kept constant
in the model.\footnote{SNR shocks propagating into a
dense medium with solar metallicity can be radiative. \citet{hollenbach79}
give the cooling length of shock heated gas (as post shock column density) by
using the cooling function of~\citet{raymond76}, $N_{\rm
c}\simeq2\times10^{17}\beta^{0.6}V_{\rm sh,7}{}^{4.2}~{\rm cm^{-2}}$, where
$V_{{\rm sh},m}\equiv10^m~{\rm cm~s^{-1}}$. Here we regard the cooling
function as determined by the electron temperature. In this paper, we
consider $V_{\rm sh}\ga10^8~{\rm cm~s^{-1}}$ for which $N_{\rm c}\ga
10^{21.2}~{\rm cm^{-2}}$. Thus, the adiabatic shock approximation can be
valid for the atomic transition layer from $z=0$ to $z=z_{\rm in}$ at which
the column density $\sim4\times10^{16}~{\rm cm^{-2}}(n/1~{\rm cm^{-3}})$ is
lower than $N_{\rm c}$.} In the following, we neglect
the radiative recombination rate~$\sim10^{-13}~{\rm cm^3~s^{-1}}$, which is
much smaller than any other rates. Moreover, we assume $n_{{\rm H},j\ne{\rm
1s}}\ll n_{\rm H,1s}$~(see Eq.~\eqref{eq:2s simple}). We will address these
assumptions later. Then, the spatial distribution of narrow hydrogen atoms is
given by
%
\begin{eqnarray}
\frac{\upartial n_{\rm H,1s}^{\rm N} }{\upartial z}
= -n_{\rm H,1s}^{\rm N}\frac{ C_{\rm I,N} + C_{\rm CX,N}}{V_{\rm sh}},
\label{eq:rate narrow}
\end{eqnarray}
%
where we define the collisional ionization rate,
%
\begin{eqnarray}
C_{\rm I,N}&=&\sum_{q=\{{\rm e,p}\}} n_q \int f_{\rm H}^{\rm N}f_q
\Delta v_q\sigma_q^{\rm I}
{\rm d}^3\bm{v_{\rm H}}
{\rm d}^3\bm{v_q},
\label{eq:I narrow}
\end{eqnarray}
%
and the charge-exchange rate,
%
\begin{eqnarray}
C_{\rm CX,N}&=& n_{\rm p} \int f_{\rm H}^{\rm N}f_{\rm p}
\Delta v_{\rm p}\sigma_{\rm CX}
{\rm d}^3\bm{v_{\rm H}}
{\rm d}^3\bm{v_{\rm p}}.
\label{eq:CX naroow}
\end{eqnarray}
%
Here $n_q$ is the number density of particle $q$, the symbols $\bm{v_q}$ and
$\bm{v}_{\rm H}$ denote the velocity vectors of particle $q$ and the hydrogen
atom, respectively, $\Delta v_q\equiv\big|\bm{v_{\rm H}}-\bm{v_q}\big|$ is
the relative velocity between the hydrogen atom and particle $q$,
$\sigma_q^{\rm I}$ is the ionization cross-section by collision with particle
$q$ and $\sigma_{\rm CX}$ is the total cross-section of charge-exchange
reactions. We omit the rates of collisions between
hydrogen atoms, which are small compared with the rates by proton/electron
collisions because of the lack of shock compression or heating. Note that the
collisional rates of He$^{2+}$ impacts, which are also omitted in this paper,
would make a moderate contribution~\citep{laming96}. We assume the velocity
distribution function of narrow hydrogen atoms to be
%
\begin{eqnarray}
f_{{\rm H}}^{\rm N}
&=&
\left( \frac{m_{\rm H}}{2\upi k_{\rm B}T_0} \right)^{\frac{3}{2}}
\exp\left[
-\frac{ m_{\rm H}\left( \bm{v_{\rm H}} - \bm{V_{\rm sh}} \right)^2 }{2k_{\rm B}T_0}
\right],
\end{eqnarray}
%
where $m_{\rm H}$ is the hydrogen atom mass and $\bm{V_{\rm sh}}=(0,0,V_{\rm
sh})$. Similarly, the distribution functions of the downstream
protons/electrons are given by
%
\begin{eqnarray}
f_q
&=& \left( \frac{m_q}{2\upi k_{\rm B}T_q} \right)^{\frac{3}{2}}
\exp\left[
-\frac{ m_q\left( \bm{v_q} - \bm{u_2} \right)^2 }{2k_{\rm B}T_q}
\right],
\end{eqnarray}
%
where $\bm{u_2}=(0,0,u_2)$ and $q=\{\rm p,e \}$.
For simplicity, we assume that the broad
atoms have the same mean velocity and temperature as the downstream protons.
Note that the velocity distribution of broad atoms can substantially deviate from
the proton's due to the velocity dependence of the cross-section of charge-exchange
reaction~\citep[e.g.][]{heng07},
but the nature of radiative line transfer
depends mainly on the ionization structure of `narrow' atoms, which can be well
approximated by
Eq.~\eqref{eq:rate narrow}.
Then, for the broad hydrogen atoms, we obtain the
distribution function,
%
\begin{eqnarray}
f_{{\rm H}}^{\rm B}
&=&
\left( \frac{m_{\rm H}}{2\upi k_{\rm B}T_{\rm p}} \right)^{\frac{3}{2}}
\exp\left[
-\frac{ m_{\rm H}\left( \bm{v_{\rm H}} - \bm{u_2} \right)^2 }{2k_{\rm B}T_{\rm p}}
\right],
\end{eqnarray}
%
and the differential equation of their spatial distribution,
%
\begin{eqnarray}
\frac{\upartial n_{\rm H,1s}^{\rm B} }{\upartial z}
= \frac{n_{\rm H,1s}^{\rm N}C_{\rm CX,N}
       -n_{\rm H,1s}^{\rm B}C_{\rm I,B}
       }{u_2},
\label{eq:rate broad}
\end{eqnarray}
%
where
%
\begin{eqnarray}
C_{\rm I,B}&=&\sum_{q=\{{\rm e,p}\}} n_q \int f_{\rm H}^{\rm B}f_q
\Delta v_q\sigma_q^{\rm I}
{\rm d}^3\bm{v_{\rm H}}
{\rm d}^3\bm{v_q}.
\label{eq:I braod}
\end{eqnarray}
%
Accordingly, the differential equation for the spatial distribution of
downstream protons is
%
\begin{eqnarray}
\frac{\upartial n_{\rm p}}{\upartial z}
= \frac{n_{\rm H,1s}^{\rm N}C_{\rm I,N}+n_{\rm H,1s}^{\rm B}C_{\rm I,B}}{u_2}.
\label{eq:rate proton}
\end{eqnarray}
%
Note that the electron number density is given by
the charge neutrality condition $n_{\rm e}=n_{\rm p}$.
We consider the radiative line transfer and the population of
bound-state~($j\neq{\rm 1s}$) atoms under the ionization structure
given by the above formulae.
\par
Here we consider the population of excited hydrogen atoms.
The rate equation for the excited-state $j$ is
%
\begin{eqnarray}
\frac{{\rm d} n_{{\rm H},j}}{{\rm d}t}=\sum_k
\left\{
 n_{{\rm H},k}\left( C_{k,j}+P_{k,j} \right)
-n_{{\rm H},j}\left( C_{j,k}+P_{j,k} \right)
\right\},
\label{eq:rate}
\end{eqnarray}
%
where $P_{k,j}$ and $C_{k,j}$
($P_{j,k}$ and $C_{j,k}$)
are radiative and collisional rates per unit time
for the transition from $k$ to $j$ ($j$ to $k$), respectively.
The collisional rate is
%
\begin{eqnarray}
C_{j,k}=\sum_{q=\{{\rm e,p}\}}n_q
\int f_{{\rm H}}(\bm{v_{\rm H}})
f_q(\bm{v_q})
\Delta v_q
\sigma_q^{j,k}
{\rm d}^3\bm{v_{\rm H}}
{\rm d}^3\bm{v_q},
\label{eq:collisional rate}
\end{eqnarray}
%
where $\sigma_q^{k,j}$ is a sum of all kinds of collisional cross-sections
between particle $q$ and hydrogen atom resulting in the transition from $k$
to $j$. Here we do not distinguish between the broad and narrow hydrogen
atoms so that $f_{{\rm H},j}=f_{\rm H}^{\rm N}+f_{\rm H}^{\rm B}$. For
$n_k>n_j$ (henceforth, we refer $k>j$), the radiative rates are
%
\begin{eqnarray}
P_{k,j} = A_{k,j} - \int\frac{4\upi}{h\nu} {\rm d}\nu
\int_{-1}^{1} \frac{\sigma_{\nu,\mu}^{k,j} I_{\nu,\mu}}{2} {\rm d}\mu,
\label{eq:radiative rate upper}
\end{eqnarray}
%
and
%
\begin{eqnarray}
P_{j,k} = \int\frac{4\upi}{h\nu} {\rm d}\nu
\int_{-1}^{1} \frac{\sigma_{\nu,\mu}^{j,k} I_{\nu,\mu}}{2} {\rm d}\mu,
\label{eq:radiative rate lower}
\end{eqnarray}
%
where $A_{k,j}$ is the rate of the spontaneous transition from $k$ to $j$ and
$\mu\equiv\cos\theta$ indicates the direction of ray making an angle $\theta$
to the $z$-axis (see Fig.~\ref{fig:shock}a). Here $\sigma_{\nu,\mu}^{k,j}$ is
the radiative cross-section for the ray in the direction $\mu$ at the
frequency $\nu$ resulting in the transition from $k$ to $j$, and
$I_{\nu,\mu}$ is the specific intensity of the ray directed in $\mu$ (in unit
erg cm$^{-2}$ s$^{-1}$ Hz$^{-1} $str$^{-1}$). In order to evaluate the
radiative rate $P_{j,k}$, we need to simultaneously solve the radiation
transfer equation
%
\begin{eqnarray}
\frac{{\rm d} I_{\nu,\mu}}{{\rm d}s}= -k_{\nu,\mu} I_{\nu,\mu} + j_{\nu,\mu},
\end{eqnarray}
%
where $s$ is the unit length measured along the path of ray, and
$k_{\nu,\mu}$ and $j_{\nu,\mu}$ are the absorption coefficient and emission
coefficient (or emissivity), respectively. Note that these coefficients
depend on the occupation number of the atomic states. We give their formulae
later. According to the two free-escape boundaries at $z=z_{\rm out}$ and
$z=z_{\rm in}$, we derive the formal solution of the intensity as
%
\begin{eqnarray}
I_{\nu,\mu}(\tau_{\nu,\mu}) = \int_0^{\tau_{\nu,\mu}} S_{\nu,\mu}(\tilde{\tau}_{\nu,\mu})
{\rm e}^{-\frac{\tilde{\tau}_{\nu,\mu}-\tau_{\nu,\mu}}{\mu}} \frac{{\rm d}\tilde{\tau}_{\nu,\mu}}{\mu},
\label{eq:formal solution}
\end{eqnarray}
%
where $S_{\nu,\mu}\equiv j_{\nu,\mu}/k_{\nu,\mu}$ is the source function.
The optical depth $\tau_{\nu,\mu}$ is defined as
%
\begin{eqnarray}
\tau_{\nu,\mu}=\int_{z_{\rm out}}^{z} k_{\nu,\mu} \frac{{\rm d}z}{\mu}~~~({\rm for}~\mu>0),
\label{eq:tau for positive}
\end{eqnarray}
%
and
%
\begin{eqnarray}
\tau_{\nu,\mu}=\int_{z_{\rm in}}^{z} k_{\nu,\mu} \frac{{\rm d}z}{\mu}~~~({\rm for}~\mu<0).
\label{eq:tau for negative}
\end{eqnarray}
%
We can solve the set of equations iteratively.\footnote{ We do not solve the
momentum gain of atoms due to the absorption of photons for simplicity. For
example, the velocity change of atom for each scattering of Ly~${\rm \alpha}$
is $\sim3\times10^{-3}~{\rm km~s^{-1}}$, which is much smaller than the
thermal velocity of the hydrogen atoms $\ga 10~{\rm km~s^{-1} }$.}
\par
Several of the terms of Eq.~\eqref{eq:rate} can be omitted because of the
enormous difference in time scales. The orders of magnitude of the time
scales are $t_{\rm r}\sim10^{-8}\mathchar`-10^{-1}~{\rm s}$ for the decay
time of excited atoms due to the spontaneous transition,
$t_{\rm c}\sim10^4\mathchar`-10^8~{\rm s}~(n/1~{\rm cm^{-3}})^{-1}$ for the
collisions between the hydrogen atoms and shock heated protons/electrons, and
the mean collision time for the upstream medium with temperature of $\sim10^{4}$~K
is $\sim10^{11}$~s. Note that the radiative rates due to photon absorption $P_{j,k}\big|_{j<k}$
may be at most comparable with the maximum value of collisional excitation rates
$\sim 1/t_{\rm c}$ (in the sense of the orders of magnitude estimate) because
the radiation field consists only of the line emissions. The ratio of the
excitation rate $\sim1/t_{\rm c} $ to the decay rate $\sim1/t_{\rm r}$
indicates a very small occupation number of excited state atoms $n_{{\rm
H},j\neq1s}\sim10^{-9}\mathchar`-10^{-16}n_{\rm H,1s}$ (see Eq.~\eqref{eq:2s
simple}). In addition, the recombination rate of hydrogen atoms, $\sim
10^{-14}~{\rm cm^3~s^{-1}}$, is negligibly small. Assessing the above
factors, we can ignore the collisional deexcitation, recombination and any
other collisional processes in the upstream region. Thus, for excited states
($j\neq$~1s), we obtain the statistical equilibrium condition
%
\begin{eqnarray}
n_{\rm H,1s}C_{1s,j}
+\sum_k
\left\{
 n_{{\rm H},k}P_{k,j}
-n_{{\rm H},j}P_{j,k}
\right\}=0.
\label{eq:rate j}
\end{eqnarray}
%
Here we approximate ${\rm d} n_{{\rm H},j\neq{\rm 1s}}/{\rm d}t\approx0$ and
consider direct excitation and charge-exchange as the collisional excitation
processes.
\par
We can derive the ratio of $n_{{\rm H},j}$ to $n_{\rm H,1s}$ from
Eq.~\eqref{eq:rate j} if the radiative rates $P_{j,k}$ are given. To do this,
the absorption coefficient $k_\nu$ and the emissivity $j_\nu$ are required.
In this paper, we consider only the resonant scattering, whose cross-section
is typically $\sim10^{12}$ times the Thomson cross-section. Moreover, we
neglect the coherence of all atomic (quantum) processes for the line profile
function, because its contribution is usually very small and appears at a
frequency far from the line centre (e.g. the Lorentzian wing). Furthermore,
we ignore any overlaps in the frequency among each line. This is true for the
lines we are interested in (e.g. Ly~$\alpha$, Ly~$\beta$,~Ly~$\gamma$,
H~$\alpha$, H~$\beta$, Pa~$\alpha$ and so on). Thus, we individually treat
the specific intensities, absorption coefficients and emissivities of each
line induced by the transition from $n_k$ to $n_j$. The radiative
cross-section for the ray in the direction $\mu$ resulting in the transition
from $j$ to $k$ ($j<k$) is given by
%
\begin{eqnarray}
\sigma_{\nu,\mu}^{j,k}=\upi r_{\rm e} c f_{j,k}\phi_{\nu,\mu}^{j,k},
\label{eq:radiative cross section}
\end{eqnarray}
%
where $r_{\rm e}$ is the classical electron radius, $c$ is the speed of
light, $f_{j,k}$ is the oscillator strength, and $\phi_{\nu,\mu}^{j,k}$ is
the line profile function for the transition from $j$ to $k$:
%
\begin{eqnarray}
\phi_{\nu,\mu}^{j,k}
&=&
\frac{1}{2\sqrt{\upi}\Delta\nu_{\rm D}^{\rm N}}
\frac{n_{{\rm H},j}^{\rm N}}{n_{{\rm H},j}}
\exp\left[ -\left( \frac{ \nu-\nu_{j,k,\mu}^{{\rm N}} }{\Delta\nu_{\rm D}^{\rm N}} \right)^2 \right] \nonumber \\
&+&
\frac{1}{2\sqrt{\upi}\Delta\nu_{\rm D}^{\rm B}}
\frac{n_{{\rm H},j}^{\rm B}}{n_{{\rm H},j}}
\exp\left[ -\left( \frac{ \nu-\nu_{j,k,\mu}^{\rm B} }{\Delta\nu_{\rm D}^{\rm B}} \right)^2 \right],
\label{eq:line profile function}
\end{eqnarray}
%
where the normalization condition
$\int \phi_{\nu, \mu}^{j,k} {\rm d}\nu {\rm d}\mu=1$
is satisfied.
$\Delta\nu_{\rm D}^{\rm N}$ is
the Doppler frequency for the narrow hydrogen atoms,
%
\begin{eqnarray}
\Delta\nu_{\rm D}^{\rm N}=\nu_{j,k}'\sqrt{ \frac{2k_{\rm B}T_0}{ m_{\rm H} c^2 } },
\end{eqnarray}
%
where $\nu_{j,k}'$ is the frequency at the line centre measured in the atom
rest frame, and $\Delta\nu_{\rm D}^{\rm B}$ is the Doppler frequency for the
broad hydrogen atoms,
%
\begin{eqnarray}
\Delta\nu_{\rm D}^{\rm B}=\nu_{j,k}'\sqrt{ \frac{2k_{\rm B}T_{\rm p}}{ m_{\rm H} c^2 } }.
\end{eqnarray}
%
The centroid frequencies are, respectively,
%
\begin{eqnarray}
\nu_{j,k,\mu}^{\rm N}=\nu_{j,k}'\left( 1+\frac{V_{\rm sh}}{c}\mu \right)
\end{eqnarray}
%
and
%
\begin{eqnarray}
\nu_{j,k,\mu}^{\rm B}=\nu_{j,k}'\left( 1+\frac{u_2}{c}\mu \right).
\end{eqnarray}
%
Then, the absorption coefficient for the transition from $j$ to $k$ is
%
\begin{eqnarray}
k_{\nu,\mu}^{j,k}
&=& \sigma_{\nu,\mu}^{j,k} n_{{\rm H},j} + \sigma_{\nu,\mu}^{k,j} n_{{\rm H},k}~~~({\rm for}~k>j)
\nonumber \\
&=& \upi r_{\rm e} c f_{j,k}\phi_{\nu,\mu}^{j,k}\left(n_{{\rm H},j} -\frac{g_{l_j}}{g_{l_k}}n_{{\rm H},k}\right),
\end{eqnarray}
%
where $g_{l_j}$ is the statistical weight of the state $j$ and we use the
relation $f_{k,j}=-\frac{g_{l_j}}{g_{l_k}}f_{j,k}$. Note that
$\sigma_\nu^{k,j}n_{{\rm H},k}$ indicates the stimulated emission. In the
following, we consider only dipole transitions ($| l_j - l_k|=1$) for the
absorption coefficient. The net absorption coefficient for each line
resulting in the transition from $n_j$ to $n_k$ is given by
%
\begin{eqnarray}
k_{\nu,\mu}=\sum_{j<k,k} k_{\nu,\mu}^{j,k},
\label{eq:net absp}
\end{eqnarray}
%
where for fixed $n_j$ and $n_k$, we take the summation of $k_{\nu,\mu}^{j,k}$
for $l_j$ and $l_k$ under the constraint $| l_j - l_k|=1$. For the
emissivity, we consider not only the dipole transition but also the
transition from 2s to 1s (i.e. 2$\gamma$--decay). This 2s--1s transition
yields two photons to satisfy the conservation of net angular momentum before
and after the transition. The frequencies of two photons range
$0<\nu<\nu_{\rm Ly\alpha}$ and their sum is equal to the frequency of
Ly~$\alpha$, $\nu_{\rm Ly\alpha}$, because of energy conservation. The
spontaneous 2s-1s transition rate depends on the frequency. The net
transition rate is given by
%
\begin{eqnarray}
A_{\rm 2s,1s}=\frac{1}{2}\int_{0}^{\nu_{\rm Ly\alpha}} \varphi_{\nu}
\frac{{\rm d}\nu}{\nu_{\rm Ly\alpha}},
\end{eqnarray}
%
where $\varphi_{\nu}$ is equivalently the line profile function. This has a
peak at a frequency of $\nu=0.5\nu_{\rm Ly\alpha}$ and has the half-maximum
values at $\nu\simeq0.1\nu_{\rm Ly\alpha}$ and $\nu\simeq0.9\nu_{\rm
Ly\alpha}$~\citep[e.g.][]{chluba08}. Thus, the 2$\gamma$-decay has a much
wider profile compared with $\phi_{\nu,\mu}^{j,k}$ and therefore behaves as
continuum emission with respect to each line with frequency lower than
$\nu_{\rm Ly\alpha}$ (i.e. $\varphi_\nu=0$ for $\nu>\nu_{\rm Ly\alpha}$).
Hence, the emissivity for the ray directed along $\mu$ at the frequency $\nu$
for each line is
%
\begin{eqnarray}
j_{\nu,\mu}=\sum_{k>j,j} \frac{h\nu}{4\upi} n_{{\rm H},k} A_{k,j} \phi_{\nu,\mu}^{k,j}
+\frac{h\nu}{4\upi}n_{\rm H,2s}\frac{\varphi_\nu}{2\nu_{\rm Ly\alpha}},
\label{eq:net emis}
\end{eqnarray}
%
where we take the Doppler shift into account for the line profile function
$\phi_{\nu,\mu}^{k,j}$ but neglect it for the 2$\gamma$-decay profile
$\varphi_\nu$. Note that the line profile function for the emissivity,
$\phi_{\nu,\mu}^{k,j}$, has the same form as Eq.~\eqref{eq:line profile
function} because the direction $\mu$ is fixed and any coherences are
ignored. Using the absorption coefficient $k_{\nu,\mu}$ (Eq.~\eqref{eq:net
absp}) and the emissivity $j_{\nu,\mu}$ (Eq.~\eqref{eq:net emis}), we derive
the specific intensity of each line from its formal solution
Eq.~\eqref{eq:formal solution}. Then, the radiative rates,
Eqs.~\eqref{eq:radiative rate upper} and \eqref{eq:radiative rate lower}, are
calculated. Hence, we obtain the ratio $n_{{\rm H},j}$ to $n_{\rm H,1s}$ from
Eq.~\eqref{eq:rate j}. Using the ratio of $n_{{\rm H},j}$ to $n_{\rm H,1s}$,
we obtain newly $k_{\nu,\mu}$ and $j_{\nu,\mu}$, solving the radiation field,
and eventually obtain the ratio again. We iterate the above procedure until
the ratio converges. Note that $n_{\rm H,1s}=n_{\rm H,1s}^{\rm N}+n_{\rm
H,1s}^{\rm B}$ and $n_{\rm p}$ are solved separately from any excited states
$n_{{\rm H},j\neq{\rm 1s}}$ by Eqs.~\eqref{eq:rate narrow}, ~\eqref{eq:rate
broad} and \eqref{eq:rate proton}.~\footnote{ Here we neglect the Lorentz
transformations for the intensity $I_{\nu,\mu}=(\nu/\nu')^3I'_{\nu',\mu'}$,
the emissivity $j_{\nu,\mu}=(\nu/\nu')^2j'_{\nu',\mu'}$, and the absorption
coefficient $k_{\nu,\mu}=(\nu'/\nu)k'_{\nu',\mu'}$, where prime indicates the
rest frame of hydrogen atom, i.e. $I_{\nu,\mu}\approx I'_{\nu',\mu'}$,
$j_{\nu,\mu}\approx j'_{\nu',\mu'}$ and $k_{\nu,\mu}\approx k'_{\nu',\mu'}$.
}
\par
For the numerical calculation of $n_{{\rm H},j}$, we use data for collisional
cross-sections and their fitting functions provided in the
literature~\citep[][]{barnett90,janev93,bray95,heng08,tseli12}. Note that
since the available data are limited, we calculate states up to 4f.
Therefore, we solve Ly~$\alpha$, Ly~$\beta$, Ly~$\gamma$, H~$\alpha$,
H~$\beta$ and Pa~$\alpha$. Moreover, the data of direct-collisional
excitation to $n_j=4$ level by proton impact are unavailable for the range
$v_{\rm H}\le1000~{\rm km~s^{-1}}$~\citep{tseli12}. We treat them to be zero
in this range~\citep[see also][]{shimoda18}. We also refer to the data table
for the spontaneous transition rates, the oscillator strengths and the
centroid frequencies $\nu'_{j,k}$ provided by~\citet{wiese09}.
Table~\ref{tab:A and f} presents the $A_{k,j}$ and
$f_{j,k}$ we are interested in. The centroid frequencies measured in the atom
rest frame, $\nu'_{j,k}$, are calculated from the centroid
wavelengths~\citep[$\lambda_{\rm vac}$ in][]{wiese09} divided by the speed of
light. Here we assume $\nu'_{\rm 2s,1s}=\nu'_{\rm 2p,1s}$. Note that the
offset of centroid frequencies, for example, $1-\nu'_{\rm 2s, 3p}/\nu'_{\rm
2p,3s}\simeq1.8\times10^{-5}$ is comparable with the Doppler shift of
hydrogen atoms with velocity of $\simeq 5~{\rm km~s^{-1}}$. Thus, the widths
of Balmer and Paschen lines will be broader than the width given by only the
thermal Doppler shift. Moreover, we use the fitting function of $\varphi_\nu$
provided by~\citet{chluba08}. Note that the statistical weights for each
state are $g_{\rm s}=2$, $g_{\rm p}=6$, $g_{\rm d}=10$ and $g_{\rm f}=14$,
respectively.
%
\begin{table}
\centering
\caption{Data of $A_{k,j}$, $f_{k,j}$ and $\nu_{j,k}'$
from~\citet{wiese09}. E$m$ indicates $\times10^{m}$}.
\label{tab:A and f}
\begin{tabular}{cccc} 
\hline
 $j$--$k$  & $A_{k,j}$ [s$^{-1}$] & $f_{j,k}$ & $\nu'_{j,k}$ [Hz] \\
\hline
\hline
   1s--2s  &  8.2206E0  &        N/A & 2.46607E15 \\
   1s--2p  &  6.2649E8  &  4.1641E-1 & 2.46607E15 \\
   1s--3p  &  1.6725E8  &  7.9142E-2 & 2.92275E15 \\
   1s--4p  &  6.8186E7  &  2.9006E-2 & 3.08257E15 \\
\hline
   2s--3p  &  2.2448E7  &  4.3508E-1 & 4.56684E14 \\
   2s--4p  &  9.6681E6  &  1.0282E-1 & 6.16521E14 \\
   2p--3s  &  6.3143E6  &  1.3598E-2 & 4.56676E14 \\
   2p--3d  &  6.4651E7  &  6.9615E-1 & 4.56679E14 \\
   2p--4s  &  2.5784E6  &  3.0468E-3 & 6.16514E14 \\
   2p--4d  &  2.0625E7  &  1.2186E-1 & 6.16516E14 \\
\hline
   3s--4p  &  3.0651E6  &  4.8495E-1 & 1.59839E14 \\
   3p--4s  &  1.8356E6  &  3.2270E-2 & 1.59836E14 \\
   3p--4d  &  7.0376E6  &  6.1860E-1 & 1.59838E14 \\
   3d--4p  &  3.4757E5  &  1.0999E-2 & 1.59836E14 \\
   3d--4f  &  1.3788E7  &  1.0181E0  & 1.59836E14 \\
\hline
\hline
\end{tabular}
\end{table}
%

\section{Atomic population of hydrogen atoms in supernova remnant shocks}
\label{sec:population} In this section, the results of the atomic population
calculations are exhibited. We parameterize the shock by the total number
density in the upstream region $n_{\rm tot,0}\equiv n_{\rm H,out}+n_{\rm
p,out}$, the upstream ionization degree (or the upstream proton fraction)
$\chi_0\equiv n_{\rm p,out}/n_{\rm tot,0}$, the downstream proton temperature
$T_{\rm p}$ and the electron temperature $T_{\rm e}$ (i.e. $\beta=T_{\rm
e}/T_{\rm p}$). The two escape-boundaries are set at $z_{\rm
out}=-5\times10^{16}~{\rm cm}\left(\frac{n_{\rm tot,0}}{1~{\rm
cm^{-3}}}\right)^{-1}$ and $z_{\rm in}=4u_2/(n_{\rm tot,0}C_{\rm I,B})$.
\par
\begin{figure}
\centering
\includegraphics[scale=0.9]{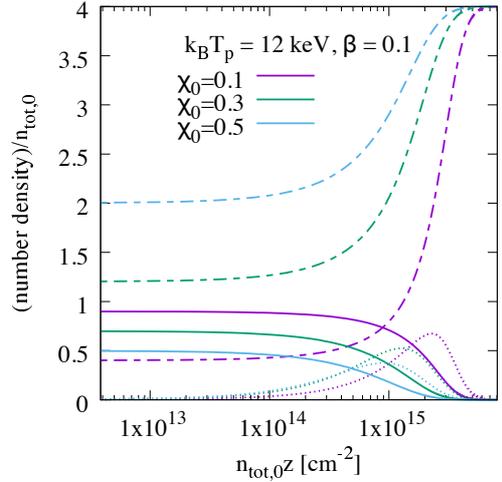}
\caption{
The downstream ionization structure for the given
$k_{\rm B}T_{\rm p}=12$~keV and $\beta=0.1$
with fixed values $\chi_0=$(0.1, 0.3 and 0.5).
The solid line, dots and broken line
show the number density of narrow hydrogen atoms,
broad hydrogen atoms and protons, respectively.
The colours correspond to $\chi_0=0.1$ (purple),
$\chi_0=0.3$ (green) and $\chi_0=0.5$ (light blue).
Note that each density fraction is self-similar function
with respect to $n_{\rm tot,0}z$.
}
\label{fig:fraction}
\end{figure}

Firstly, we give the ionization structure of hydrogen. By
rewriting Eq.~\eqref{eq:rate narrow} as
%
\begin{eqnarray}
\frac{\upartial (n_{\rm H,1s}^{\rm N}/n_{\rm tot,0}) }{\upartial (n_{\rm tot,0}z)}
= -\frac{n_{\rm H,1s}^{\rm N}}{n_{\rm tot,0}}\frac{ C_{\rm I,N} + C_{\rm CX,N}}{n_{\rm tot,0}V_{\rm sh}},
\end{eqnarray}
%
we can easily recognize that each density fraction $n_{\rm H}^{\rm N}/n_{\rm tot,0}$,
$n_{\rm H}^{\rm B}/n_{\rm tot,0}$ and $n_{\rm p}/n_{\rm tot,0}$ is
a self-similar function with respect to $n_{\rm tot,0}z$
(having the same value at the same $n_{\rm tot,0}z$)
for fixed $T_{\rm p}$, $\beta$ and $\chi_0$.
Fig.~\ref{fig:fraction} shows
$n_{\rm H,1s}^{\rm N}/n_{\rm tot,0}$,
$n_{\rm H,1s}^{\rm B}/n_{\rm tot,0}$,
and
$n_{\rm p}/n_{\rm tot,0}$
for given $k_{\rm B}T_{\rm p}=12~{\rm keV}$ and $\beta=0.1$
with fixed values $\chi_0=$(0.1, 0.3 and 0.5).
%
%
\par
The spatial distribution of $n_{{\rm H},j}$ is characterized by the ratio of
the mean free path of Lyman lines $l\propto1/n_{\rm H,1s}$ to the ionization
length of hydrogen atoms $L_{\rm I}\propto 1/n_{\rm p}$. Other series hardly
contribute to the occupation number because $n_{{\rm H},j\ne{\rm 1s}}/n_{\rm
H,1s}\ll1$. The ratio,
%
\begin{eqnarray}
\frac{l}{L_{\rm I}}\propto
\frac{n_{\rm p}}{n_{\rm H,1s}},
\end{eqnarray}
%
indicates that the spatial distribution of $n_{{\rm H},j\ne{\rm 1s}}$ has
also self-similarity with respect to $n_{\rm tot,0}z$.
In addition, from
Eq.~\eqref{eq:2s simple} or Eq.~\eqref{eq:rate j}, we recognize the scaling
relation $n_{{\rm H},j\neq{\rm 1s}}\propto n_{\rm tot,0}{}^2$. Thus, for
fixed $T_{\rm p}$, $\beta$ and $\chi_0$, $n_{{\rm H},j\neq{\rm 1s}}/n_{\rm
tot,0}{}^2$ has the same value at the same $n_{\rm tot,0}z$.
Note that this scalability of the excited state may be kept up to $n_{\rm tot,0}
\sim10^4~{\rm cm^{-3}}$ in which the occupation number of 2s state is comparable
with the ground state (i.e. our formulation would not be valid).
%
\begin{figure}
\centering
\includegraphics[scale=0.9]{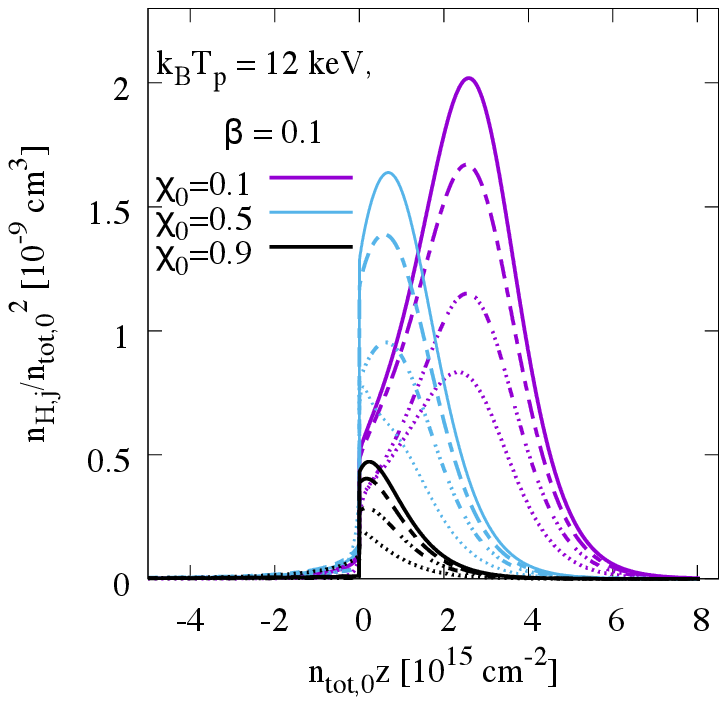}
\includegraphics[scale=0.9]{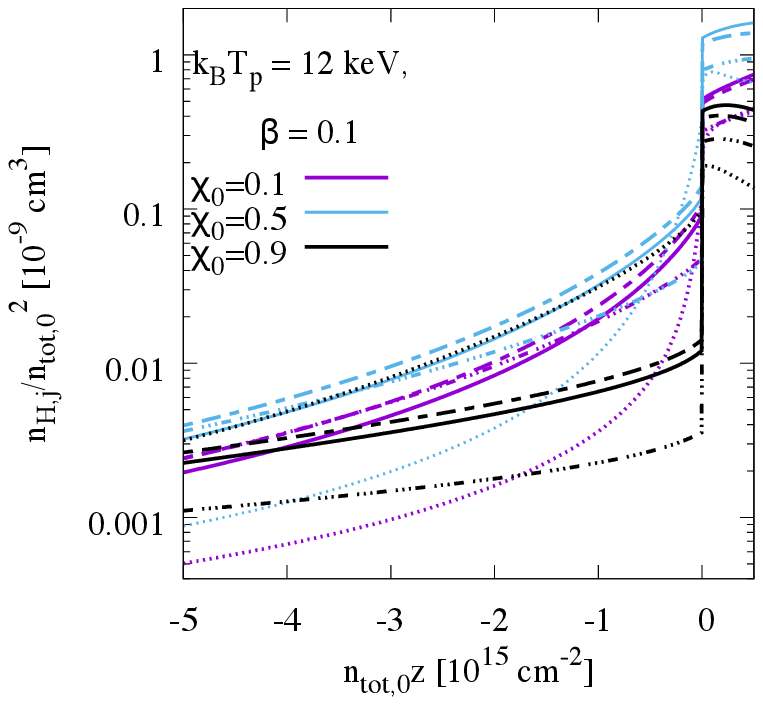}
\caption{
({\it top}):
The spatial structure of excited hydrogen atoms for given
$k_{\rm B}T_{\rm p}=12$~keV and $\beta=0.1$
with fixed values $\chi_0$=(0.1, 0.5, and 0.9).
We display $n_{\rm H,2s}$ (solid line),
$n_{\rm H,2p}\times10^{6.5}$ (dots),
$(n_{\rm H,3s}+n_{\rm H,3p}+n_{\rm H,3d})\times10^{6.5}$ (broken line)
and
$(n_{\rm H,4s}+n_{\rm H,4p}+n_{\rm H,4d}+n_{\rm H,4f})\times10^{6.5}$ (line with three dots).
The colours correspond to $\chi_0=0.1$ (purple),
$\chi_0=0.5$ (light blue) and $\chi_0=0.9$ (black).
Note that each occupation number $n_{{\rm H},j\neq{\rm 1s}}/n_{\rm tot,0}{}^2$
is a self-similar function with respect to $n_{\rm tot,0}z$.
({\it bottom}):
Close up of the upstream region.
}
\label{fig:occupation}
\end{figure}
%
%
\begin{figure}
\centering\centering
\includegraphics[scale=0.9]{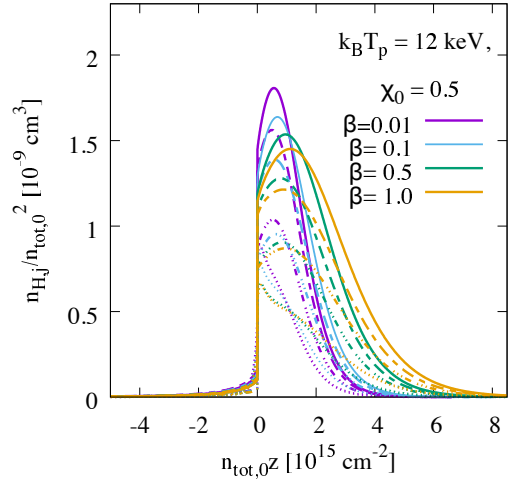}
\caption{
The spatial structure of excited hydrogen atoms for given
$k_{\rm B}T_{\rm p}=12~{\rm keV}$ and $\chi_0=0.5$
with fixed values $\beta=$(0.01, 0.1, 0.5 and 1).
We display $n_{\rm H,2s}$ (solid line),
$n_{\rm H,2p}\times10^{6.5}$ (dots),
$(n_{\rm H,3s}+n_{\rm H,3p}+n_{\rm H,3d})\times10^{6.5}$ (broken line)
and
$(n_{\rm H,4s}+n_{\rm H,4p}+n_{\rm H,4d}+n_{\rm H,4f})\times10^{6.5}$ (line with three dots).
The colours correspond to
$\beta=0.01$ (purple),
$\beta=0.1$ (light blue)
$\beta=0.5$ (green)
and $\beta=1$ (orange).
Note that each occupation number $n_{{\rm H},j\neq{\rm 1s}}/n_{\rm tot,0}{}^2$
is a self-similar function with respect to $n_{\rm tot,0}z$.
}
\label{fig:occupation-beta}
\end{figure}
%
Fig.~\ref{fig:occupation} represents the spatial distribution of $n_{{\rm
H},j}/n_{\rm tot,0}{}^2$ for given $k_{\rm B}T_{\rm p}=12~{\rm keV}$ and
$\beta=0.1$ with fixed values $\chi_0=$(0.1, 0.5 and 0.9), while
Fig.~\ref{fig:occupation-beta} represents the case of given $k_{\rm B}T_{\rm
p}=12~{\rm keV}$ and $\chi_0=0.5$ with fixed values $\beta=$(0.01, 0.1, 0.5
and 1). In this case, the maximum value of each occupation number varies
$\sim10$ per cent due to the variation of $\beta$. Note that although there
are no energetic particles leaking and no extra radiation sources (e.g.
thermal emission from the SNR ejecta), the shock-precursor-like structure is
formed by Lyman lines (see the bottom panel of Fig.~\ref{fig:occupation}).
%
\begin{figure}
\centering
\includegraphics[scale=0.9]{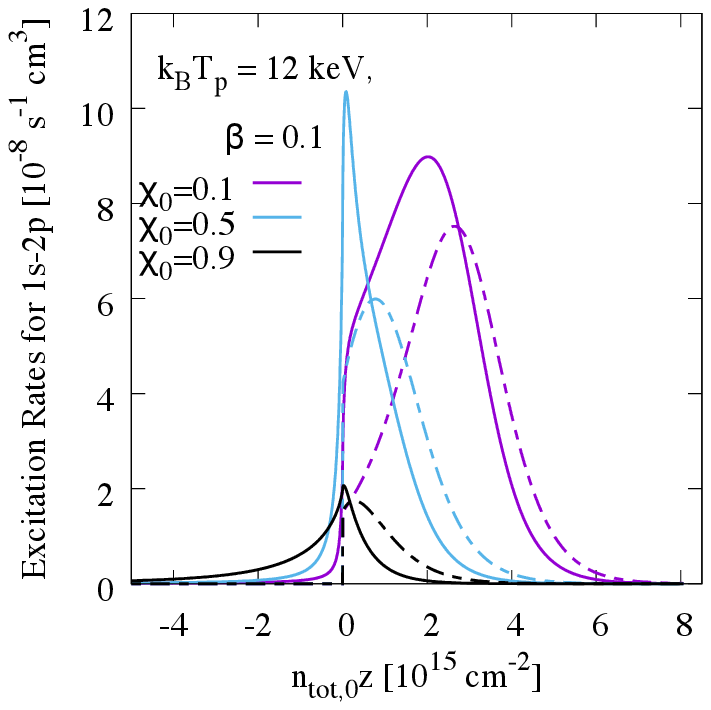}
\includegraphics[scale=0.9]{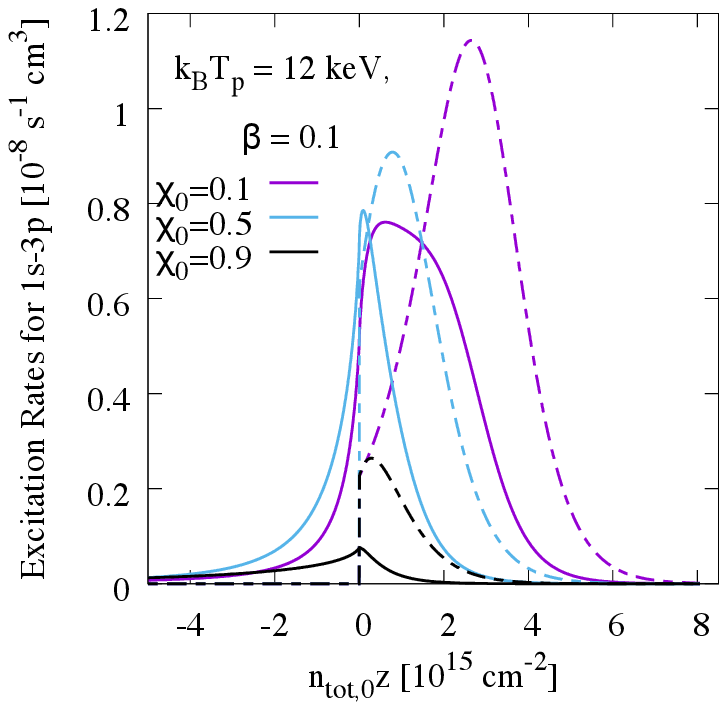}
\includegraphics[scale=0.9]{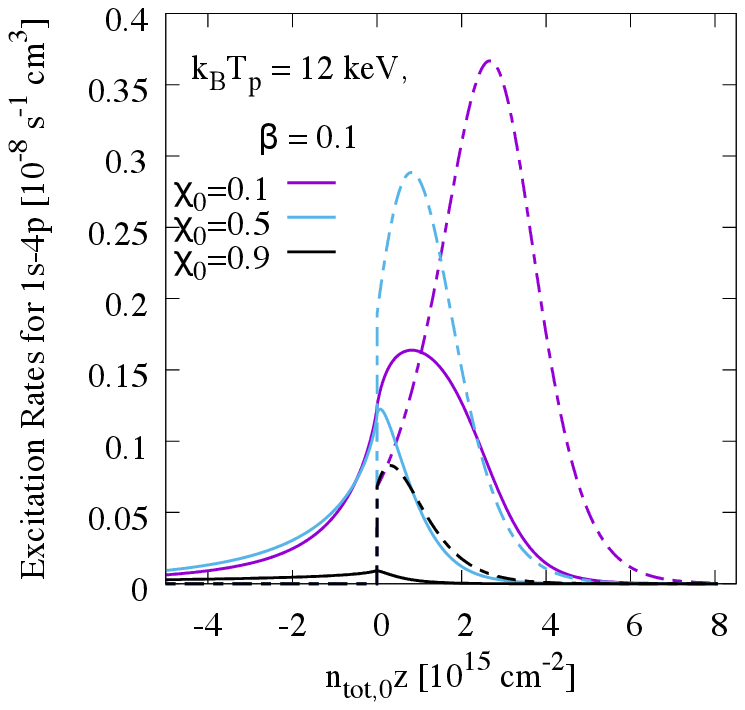}
\caption{
({\it top}):
The excitation rates for the transition from 1s to 2p
for given $k_{\rm B}T_{\rm p}=12~{\rm keV}$ and $\beta=0.1$
with fixed values $\chi_0=$(0.1, 0.5 and 0.9).
The solid line shows the radiative rate, $n_{\rm H,1s}P_{\rm 1s,2p}/n_{\rm tot,0}{}^2$,
while the broken line shows the collisional rate, $n_{\rm H,1s}C_{\rm 1s,2p}/n_{\rm tot,0}{}^2$.
The colours correspond to $\chi_0=0.1$ (purple),
$\chi_0=0.5$ (light blue) and $\chi_0=0.9$ (black).
({\it middle}):
For the transition from 2s to 3p.
({\it bottom}):
For the transition from 2s to 4p.
}
\label{fig:Lya-crpr}
\end{figure}
%
Fig.~\ref{fig:Lya-crpr}  shows the collisional excitation rates,
$n_{\rm H,1s}C_{j,k}/n_{\rm tot,0}{}^2$,
and radiative excitation rates, $n_{\rm H,1s}P_{j,k}/n_{\rm tot,0}{}^2$,
for the transitions from 1s to 2p (Ly~$\alpha$),
from 1s to 3p (Ly~$\beta$) and from 1s to 4p (Ly~$\gamma$), respectively.
The radiative rates are comparable with the collisional rates
even in the upstream region.
\par
%
\begin{figure}
\centering
\includegraphics[scale=0.9]{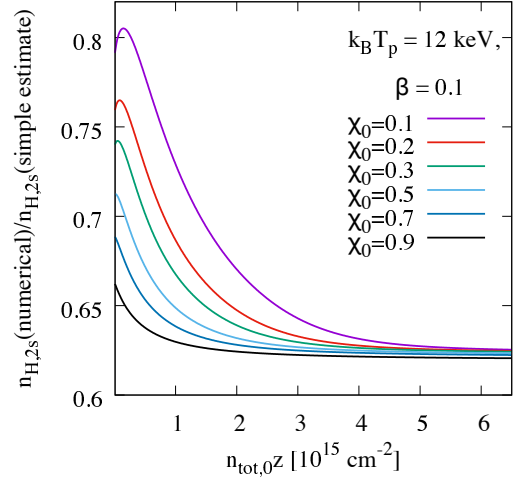}
\caption{
The ratio of numerical $n_{\rm H,2s}$ to
the simple estimate
for the optically thick limit (${\rm e}^{-\tau_0}=0$ in Eq.~\eqref{eq:2s simple}).
Here we set
$k_{\rm B}T_{\rm p}=12$~keV and $\beta=0.1$.
The colours correspond to
$\chi_0=0.1$ (purple),
$\chi_0=0.2$ (red),
$\chi_0=0.3$ (green),
$\chi_0=0.5$ (light blue),
$\chi_0=0.7$ (blue)
and $\chi_0=0.9$ (black).
}
\label{fig:simple}
\end{figure}
\begin{figure}
\centering
\includegraphics[scale=0.9]{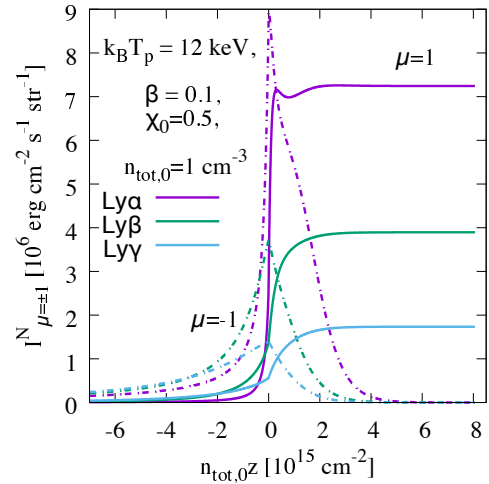}
\caption{
The intensities of
narrow Ly~$\alpha$ (purple), Ly~$\beta$ (green) and Ly~$\gamma$ (light blue)
in the directions $\mu=1$ (solid line) and $\mu=-1$ (broken line).
Here we set $k_{\rm B}T_{\rm p}=12$~keV, $\beta=0.1$, $\chi_0=0.5$ and
$n_{\rm tot,0}=1~{\rm cm^{-3}}$.
}
\label{fig:Ly-intensity}
\end{figure}

%
Fig.~\ref{fig:simple} shows a comparison of $n_{\rm H,2s}$ from numerical
calculation and $n_{\rm H,2s}$ from the simple estimate for the optically
thick limit (${\rm e}^{-\tau_0}=0$ in Eq.~\eqref{eq:2s simple}). The
numerical value of $n_{\rm H,2s}$ less than the estimated $n_{\rm H,2s}$ in
the optically thick limit implies the incomplete conversion of Lyman lines to
Balmer lines. Fig.~\ref{fig:Ly-intensity} represents intensities of narrow
Ly~$\alpha$, Ly~$\beta$ and Ly~$\gamma$ in the direction $\mu=\pm1$,
%
\begin{eqnarray}
I_{\mu=\pm 1}^{\rm N}(z)=
\int_{|\nu|<|\varpi_{\pm1}^{\rm N}|}
I_{\rm \nu,\pm 1}(z) {\rm d}\nu,
\end{eqnarray}
%
where
%
\begin{eqnarray}
\varpi_{\mu}^{\rm N}=\nu_{j,k}'\frac{\mu V_{\rm sh}-25~{\rm km~s^{-1}}}{c}.
\end{eqnarray}
%
%
%
Lyman photons escape toward the downstream region ($\mu=1$) from the shock.
We define the fraction of escaping photons as
%
\begin{eqnarray}
\varepsilon_{\rm esc}^{\rm N}\equiv
\frac{|F^{\rm N}(z_{\rm out})|+|F^{\rm N}(z_{\rm in})|}
{2\upi\int_{z_{\rm out}}^{z_{\rm in}} {\rm d}z
\int_{-1}^{1} {\rm d}\mu
\int_{|\nu|<|\varpi_{\mu}^{\rm N}|} j_{\nu,\mu}(z){\rm d}\nu},
\end{eqnarray}
%
where
%
\begin{eqnarray}
F^{\rm N}(z)=2\upi\int_{-1}^{1} I_{\mu}^{\rm N}(z)\mu {\rm d}\mu,
\end{eqnarray}
%
is the energy flux in the narrow line. Note that the sum of $|F^{\rm
N}(z_{\rm out})|$ and $|F^{\rm N}(z_{\rm in})|$ indicates the net energy
taken away from the shock by the lines.
%
\begin{figure}
\centering
\includegraphics[scale=0.9]{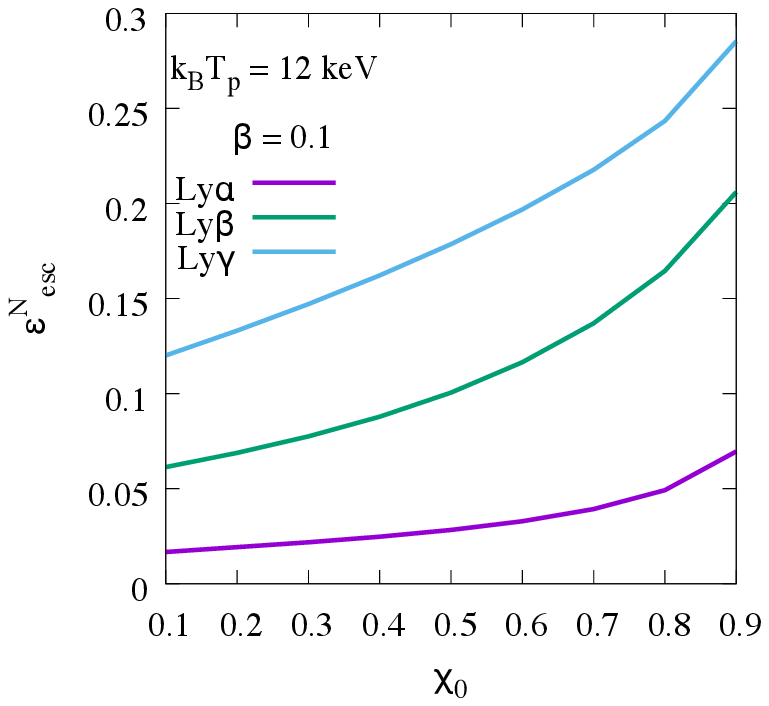}
\includegraphics[scale=0.9]{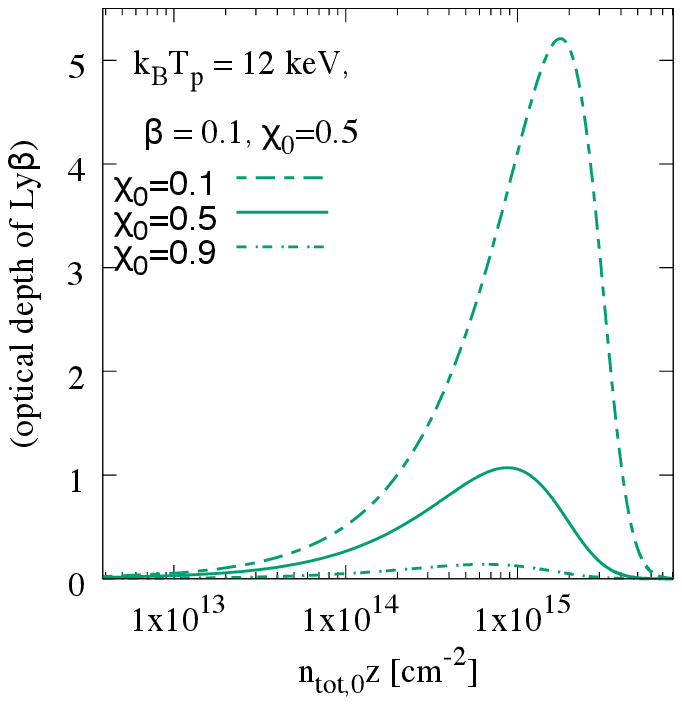}
\caption{{\it (top):}
The escape fraction of Lyman lines $\varepsilon_{\rm esc}^{\rm N}$
for the shock with $k_{\rm B}T_{\rm p}=12~{\rm keV}$ and $\beta=0.1$.
The colours represent Ly~$\alpha$ (purple), Ly~$\beta$ (green)
and Ly~$\gamma$ (light blue).
{\it (bottom):} The optical depth of Ly~$\beta$ at the line centre
measured from the shock front ($z=0$) toward downstream ($\mu=1$). Note that the optical depths of Ly~$\alpha$
and Ly~$\gamma$ are $(f_{\rm 1s,2p}/f_{\rm 1s,3p})\simeq5.26$ and $(f_{\rm 1s,4p}/
f_{\rm 1s,3p})\simeq0.367$ times the Ly~$\beta$ optical depth, respectively.
}
\label{fig:trap}
\end{figure}
%
The top panel of Fig.~\ref{fig:trap} shows the escape
fraction of the Lyman lines for a shock with $k_{\rm B}T_{\rm p}$ and
$\beta=0.1$. The fraction of Ly~$\beta$ depends somewhat
strongly on the ionization degree $\chi_0$ compared to the others. This is
because the difference in optical thickness. The bottom panel of
Fig.~\ref{fig:trap} shows the optical depth of Ly~$\beta$ at the line centre
in the direction of $\mu=1$. The depths of Ly~$\alpha$ and Ly~$\gamma$ are
$(f_{\rm 1s,2p} /f_{\rm 1s,3p})\simeq5.26$ and $(f_{\rm 1s,4p}/f_{\rm
1s,3p})\simeq0.367$ times the Ly~$\beta$ depth, respectively. The value of
Ly~$\beta$ optical-thickness decreases from $\simeq5.2$ to $\simeq0.14$ with
increasing of $\chi_0$, while the thickness of Ly~$\alpha$ and Ly~$\gamma$
vary from $\simeq27$ to $\simeq0.74$ and from $\simeq1.9$ to $\simeq0.05$,
respectively. The variations of attenuation, $e^{-\tau}$, are from
$5.5\times10^{-3}$ to $0.87$ for Ly~$\beta$, from $1.9\times10^{-12}$ to
$0.47$ for Ly~$\alpha$, and from $0.15$ to $0.95$ for Ly~$\gamma$. Thus, the
Ly~$\beta$ attenuation varies the most drastically, giving the strongest
dependence of the escape fraction on $\chi_0$. Note that the fraction would
substantially increase roughly as the velocity width of narrow hydrogen atoms
$\propto T_0{}^{\frac{1}{2}}$ because the optical depth is function of
$T_0{}^{-\frac{1}{2}}$.
Thus, the conversion of narrow Lyman lines to narrow Balmer lines is {\it incomplete}.
\par
%
\begin{figure}
\centering
\includegraphics[scale=0.9]{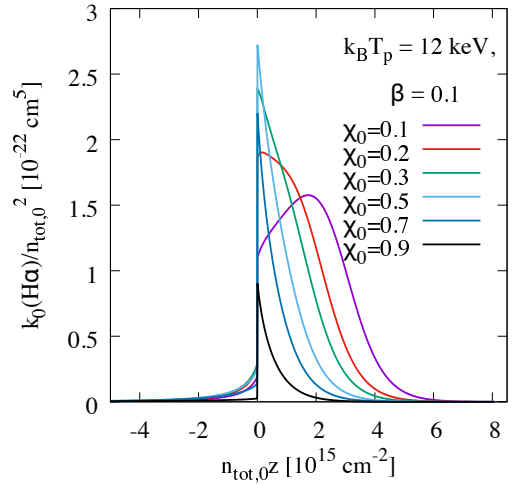}
\caption{
The absorption coefficient of H~$\alpha$ at the line centre
of the ray directed in $\mu=0.01$
for given
$k_{\rm B}T_{\rm p}=12$~keV and $\beta=0.1$
with fixed values $\chi=$( 0.1, 0.2, 0.3, 0.5, 0.7 and 0.9).
The colours correspond to
$\chi_0=0.1$ (purple),
$\chi_0=0.2$ (red),
$\chi_0=0.3$ (green),
$\chi_0=0.5$ (light blue),
$\chi_0=0.7$ (blue)
and $\chi_0=0.9$ (black).
Note that the absorption coefficient proportional to
the occupation number,
therefore it is also a self-similar function with respect to
$n_{\rm tot,0}z$ and is proportional to $n_{\rm tot,0}{}^2$.
}
\label{fig:Ha-absp}
\end{figure}
%
Fig.~\ref{fig:Ha-absp} represents the absorption coefficient of H~$\alpha$ at
the line centre of the ray directed along $\mu=0.01$ for given $k_{\rm
B}T_{\rm p}=12~{\rm keV}$ and $\beta=0.1$ with fixed values $\chi=$( 0.1,
0.2, 0.3, 0.5, 0.7 and 0.9). The value of the coefficient ranges from
$10^{-23}$ to $10^{-22}~{\rm cm^{-1}}$, which is consistent with our simple
estimate using Eq.~\eqref{eq:absp simple}. Note that the absorption
coefficient of H~$\beta$ is about quarter of $k_{\nu,\mu}({\rm H\alpha})$;
$k_{\nu,\mu}({\rm H\beta})\simeq (f_{\rm 2s,4p}/f_{\rm
2s,3p})k_{\nu,\mu}({\rm H\alpha})$.

\section{synthetic observation}
\label{sec:observation} Here we consider how lines are observed at an SNR
shock based on the calculated atomic populations. We suppose that the SNR
shock propagates into a realistic ISM, which consists of diffuse gas and
clumpy gas~\citep[e.g.][]{heiles03}, and the interaction between the shock
and the inhomogeneous medium results in H~$\alpha$ filaments on the sky
\citep[e.g.][see also Fig.~10 of \citealt{inoue12}]{hester87,shimoda15}.
Then, we presume that the shock becomes partly plane-parallel due to the
density contrast, and that the length of the sides of the plane corresponds
to the typical length scale of the density contrast. Note that the length
scale may be at least $10$ per cent of the SNR radius as indicated by the
rippling of the observed H~$\alpha$ filaments. Moreover, we regard each plane
as isolated for simplicity. In the following, our line of sight is fixed
orthogonally to the shock normal (i.e. to along the $y$-axis and $\mu=0$).

\par
Let $L_{\rm cl}$ be the extent of shock along our line of sight.
Thus, the observed intensity (surface intensity) of each line is written as
%
\begin{eqnarray}
I_{\nu,0}(z)=S_{\nu,0}(z)\left(1-{\rm e}^{-k_{\nu,0}(z)L_{\rm cl}}\right).
\end{eqnarray}
%
Note that roughly speaking, the source function hardly depends on the
frequency but the absorption coefficient has sharp peak around the line
centre. Therefore, for the optically thick case, the line shape is somewhat
flattened from the centroid frequency to the critical frequency at which
$k_{\nu,0}L_{\rm cl}\sim1$, that is, $I_{\nu,0}\approx
S_{\nu,0}=j_{\nu,0}/k_{\nu,0}$. For the frequency corresponding to
$k_{\nu,0}L_{\rm cl}\la1$, i.e. in the optically thin case, the line shape
follows the line profile function; $I_{\nu,0}\approx S_{\nu,0}k_{\nu,0}L_{\rm
cl}$. Note that we have the scaling relations $S_{\nu,0}\propto~n_{\rm
tot,0}$ for the Lyman lines, $S_{\nu,0}\propto~n_{\rm tot,0}{}^0$ for the
Balmer/Paschen lines, and $S_{\nu,0}k_{\nu,0}=j_{\nu,0}\propto n_{\rm
tot,0}{}^2$. In the following, we fix $L_{\rm cl}=2~{\rm pc}$.
Note that for SN~1006, \citet{raymond07} estimated the
ambient density $n_{\rm tot,0}=0.25\mathchar`-0.4~{\rm cm^{-3}}$, the
ionization degree $\chi_0\sim0.1$ and the length of line of sight
$\sim2\times10^{18}~{\rm cm}$, thus the optically thin condition is implied.
It is possibly consistent with the observed width of narrow component at
SN~1006, $\simeq21~{\rm km~s^{-1}}$, that is exceptionally narrow compared to
the other SNRs~\citep{sollerman03}.
\par
%
\begin{figure}
\centering
\includegraphics[scale=0.9]{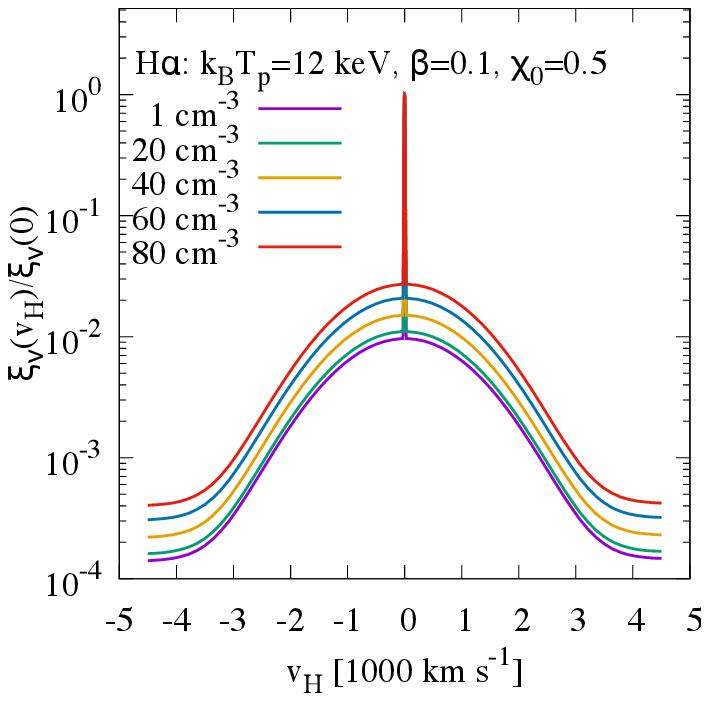}
\includegraphics[scale=0.9]{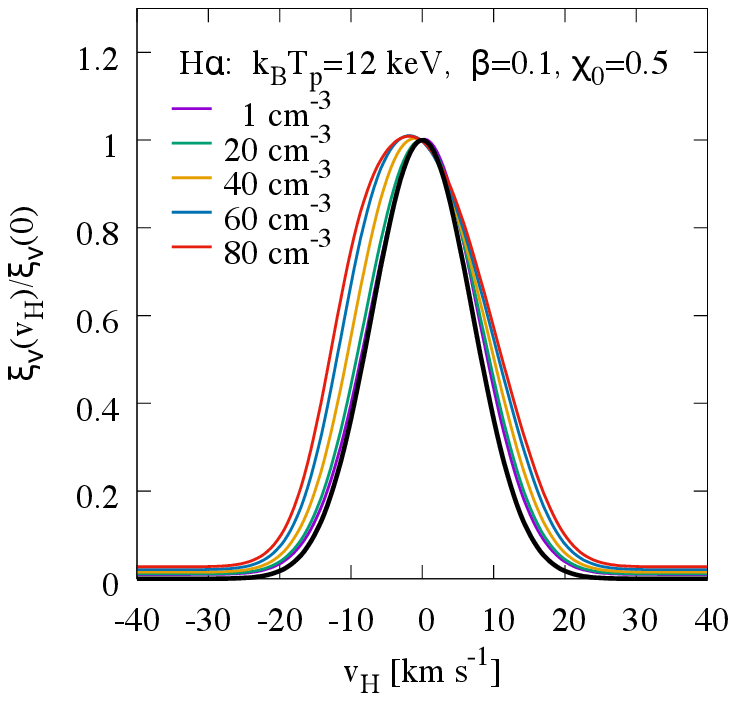}
\caption{
({\it top}):
The spectra of observed H~$\alpha$ in velocity space
for given $k_{\rm B}T_{\rm p}=12$~keV, $\beta=0.1$ and $\chi_0=0.5$.
The colours correspond to
$n_{\rm tot,0}=1~{\rm cm^{-3}}$ (purple),
$n_{\rm tot,0}=20~{\rm cm^{-3}}$ (green),
$n_{\rm tot,0}=40~{\rm cm^{-3}}$ (orange),
$n_{\rm tot,0}=60~{\rm cm^{-3}}$ (blue),
and $n_{\rm tot,0}=80~{\rm cm^{-3}}$ (red).
({\it bottom}):
Close up of the narrow component.
The black line represents the Maxwell distribution function
what we assume for the narrow hydrogen atoms.
}
\label{fig:whole-Ha}
\end{figure}
%
The observed H~$\alpha$ spectra,
%
\begin{eqnarray}
\xi_\nu = \int_{z_{\rm out}}^{z_{\rm in}} I_{\nu,0}(z) {\rm d}z,
\label{eq:xi}
\end{eqnarray}
%
are displayed in Fig.~\ref{fig:whole-Ha} in hydrogen
velocity space for given $k_{\rm B}T_{\rm p}=12$~keV and $\beta=0.1$ with
fixed values $n_{\rm tot,0}=$(1~cm$^{-3}$, 20~cm$^{-3}$, 40~cm$^{-3}$,
60~cm$^{-3}$ and 80~cm$^{-3}$). Note that we take the
interval of spatial integration from $z_{\rm out}$ to $z_{\rm in}$ for
simplicity. For shocks propagating into a tenuous medium ($\sim0.1~{\rm
cm^{-3}}$), the length of the precursor-like emission of H~$\alpha$ becomes
several times $10^{16}~{\rm cm}$ (see Figs.~\ref{fig:Lya-crpr} or
\ref{fig:Ly-intensity}). Therefore, in typical spectroscopic observations of
Galactic SNRs with a slit width of $\simeq1~{\rm arcsec}\simeq10^{16}~{\rm
cm}(d/1~{\rm kpc})$, the photo-precursor emissions would be missed. Here we
define the line centre ($v_{\rm H}=0$) by the mean value of relevant centroid
frequencies $\nu_{j,k}'$, $\nu_{\rm centre}= (\nu'_{\rm 2s,3p}+\nu'_{\rm
2p,3s}+\nu'_{\rm 2p,3d})/3\simeq4.56680\times10^{14}~{\rm Hz}$.
The two--photon emission dominates at velocity $|v_{\rm
H}|\ga3000~{\rm km~s^{-1}}$ in this case. Note that in reality, the broad
component would be becoming fainter at $|v_{\rm H}|\ga2000~{\rm km~s^{-1}}$
because of the velocity dependence of the charge-exchange cross-section, that
is not accounted for in our model. The bottom panel of
Fig.~\ref{fig:whole-Ha} shows the narrow component. The line shape evolves
asymmetrically with increasing $n_{\rm tot,0}$. This evolution is enhanced
(suppressed) for the case of lower (higher) $\chi_0$. Fig.~\ref{fig:narrow-Ha
chi} shows the cases of $\chi_0=0.1$ and $\chi_0=0.9$.
%
\begin{figure}
\centering
\includegraphics[scale=0.9]{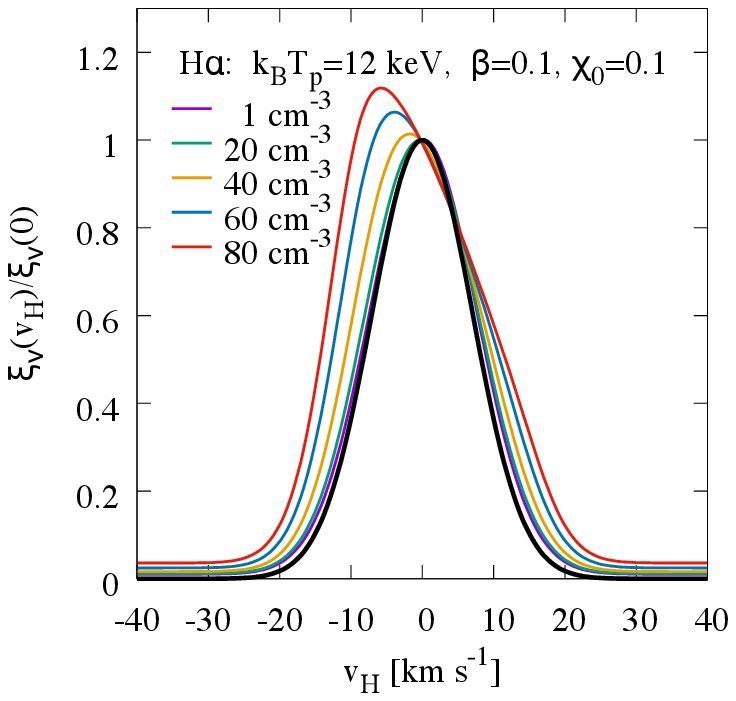}
\includegraphics[scale=0.9]{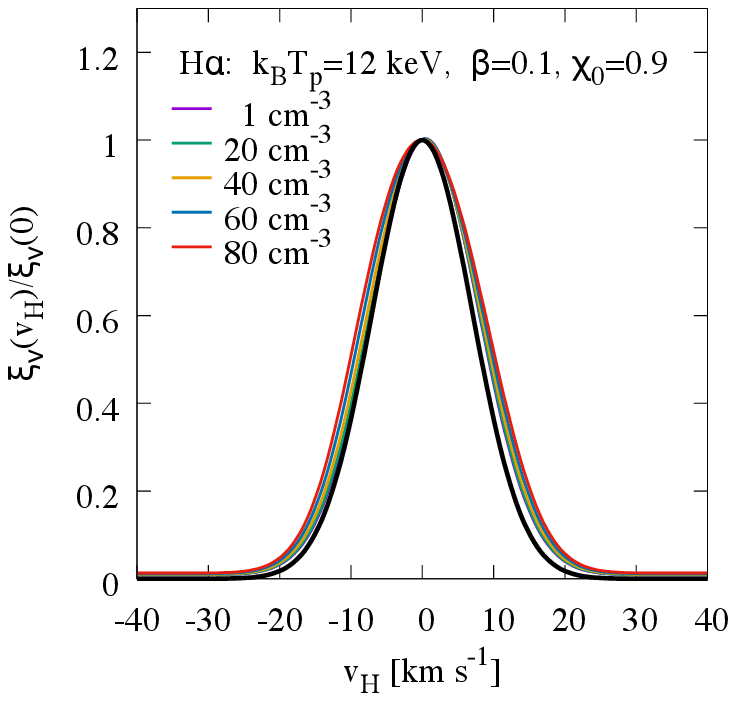}
\caption{
The kinetic spectra of narrow H~$\alpha$
for given $\chi_0=0.1$ ({\it top}) and $\chi_0=0.9$ ({\it bottom}).
$k_{\rm B}T_{\rm p}=12$~keV, $\beta=0.1$ and $\chi_0=0.5$ are also given.
The colours correspond to
$n_{\rm tot,0}=1~{\rm cm^{-3}}$ (purple),
$n_{\rm tot,0}=20~{\rm cm^{-3}}$ (green),
$n_{\rm tot,0}=40~{\rm cm^{-3}}$ (orange),
$n_{\rm tot,0}=60~{\rm cm^{-3}}$ (blue),
and $n_{\rm tot,0}=80~{\rm cm^{-3}}$ (red).
The black line represents the Maxwellian distribution function
that we assume for the narrow hydrogen atoms.
}
\label{fig:narrow-Ha chi}
\end{figure}
%
\par
These modifications of the line profile from the distribution function of
hydrogen atom come from the existence of the 2s-state hydrogen atoms and the
radiation transfer effects. The asymmetry of the line profile arises from the
offset of centroid frequencies, for instance, $1-\nu_{\rm 2s,3p}'/\nu_{\rm
centre}\simeq-0.8\times10^{-5}$ (equivalently, $v_{\rm H}\simeq+2.3~{\rm
km~s^{-1}}$).
%
\begin{figure}
\centering
\includegraphics[scale=0.9]{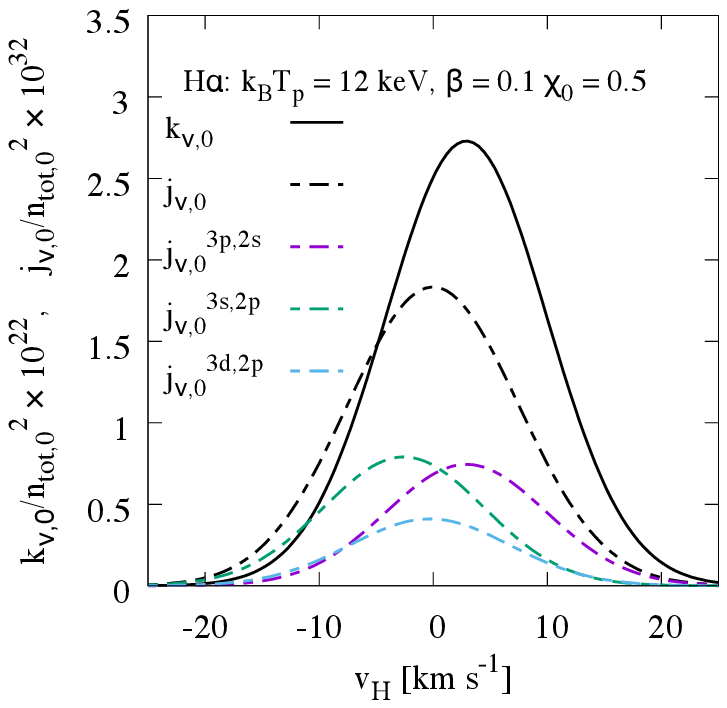}
\includegraphics[scale=0.9]{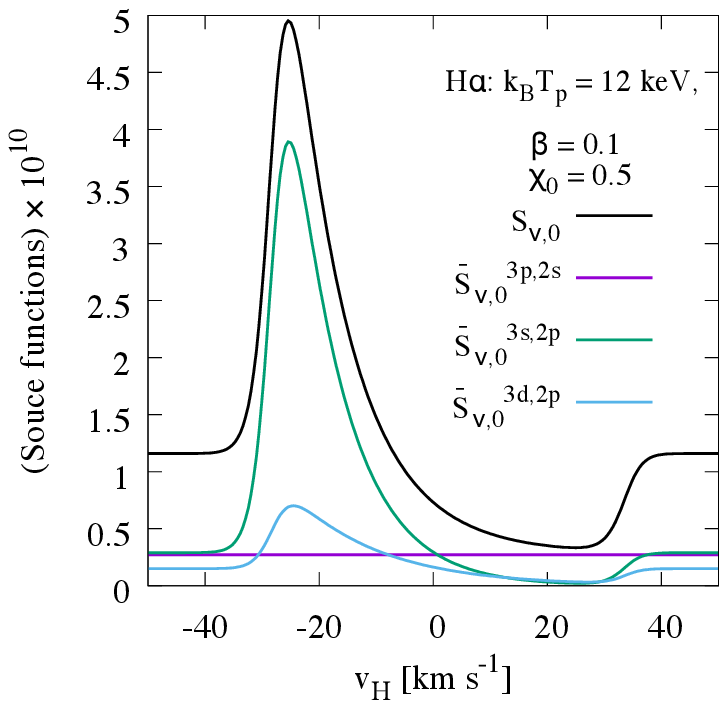}
\caption{
({\it top}): The absorption and emission coefficients of H~$\alpha$ just behind the shock front
for given $k_{\rm B}T_{\rm p}=12$~keV, $\beta=0.1$, $\chi_0=0.5$ and $n_{\rm tot}=1~{\rm cm^{-3}}$.
The solid line is the absorption coefficient,
while the broken lines are the emission coefficients.
For the emission coefficients, we display
$j_{\nu,0}$ (black),
$j_{\nu,0}^{\rm 3p,2s}$ (purple),
$j_{\nu,0}^{\rm 3s,2p}$ (green),
and
$j_{\nu,0}^{\rm 3d,2p}$ (light blue).
({\it bottom}): The source functions of H~$\alpha$ just behind the shock front
for given $k_{\rm B}T_{\rm p}=12$~keV, $\beta=0.1$ and $\chi_0=0.5$.
The solid black line shows $S_{\nu,0}$, which corresponds to the line profile in the optically thick limit.
The purple, green and light blue lines are
$\bar{S}_{\nu,0}^{\rm 3p,2s}\equiv j_{\nu,0}^{\rm 3p,2s}/k_{\nu,0}$,
$\bar{S}_{\nu,0}^{\rm 3s,2p}\equiv j_{\nu,0}^{\rm 3s,2p}/k_{\nu,0}$,
and
$\bar{S}_{\nu,0}^{\rm 3d,2p}\equiv j_{\nu,0}^{\rm 3d,2p}/k_{\nu,0}$, respectively,
where $j_{\nu,\mu}^{k,j}$ is the emissivity of each transition from $k$ to $j$
(excepted the 2$\gamma$-continuum).
Note that $v_{\rm H}=0$ corresponds to the frequency $\nu=\nu_{\rm centre}
=(\nu'_{\rm 2s,3p}+\nu'_{\rm 2p,3s}+\nu'_{\rm 2p,3d})/3$.
}
\label{fig:jnu0-Ha}
\end{figure}
%
The top panel of Fig.~\ref{fig:jnu0-Ha} shows the absorption and emission
coefficients of H~$\alpha$ just behind the shock front for a given $k_{\rm
B}T_{\rm p}=12$~keV, $\beta=0.1$, $\chi_0=0.5$ and $n_{\rm tot}=1~{\rm
cm^{-3}}$. The absorption coefficient has a peak at $v_{\rm H}\simeq2.3~{\rm
km~s^{-1}}$ (i.e. $\nu\simeq\nu'_{\rm 2s,3p}$). It is obvious because almost
all of the H~$\alpha$ is scattered by the 2s-state hydrogen atoms. On the
other hand, the emissivity $j_{\nu,0}$ has the peak at $v_{\rm H}\simeq0$
($\nu\simeq\nu_{\rm centre}$). Note that the value of peak frequency depends
on the relative occupation number among 3s, 3p and 3d states. Thus, with
increasing the optical depth, the line shape becomes more asymmetric due to
the efficient scattering around $v_{\rm H}\simeq2.3~{\rm km~s^{-1}}$, which
is not aligned with the peak frequency of emissivity. In the optically thick
limit, the line shape follows the source function that is shown in the bottom
panel of Fig~\ref{fig:jnu0-Ha}. The asymmetry of source function comes from
the offset between the peak frequencies of $j_{\nu,0}$ and $k_{\nu,0}$. Note
that as a consequence of the modification of line profile, the line width is
apparently broadened.
\par
\begin{figure}
\centering
\includegraphics[scale=0.9]{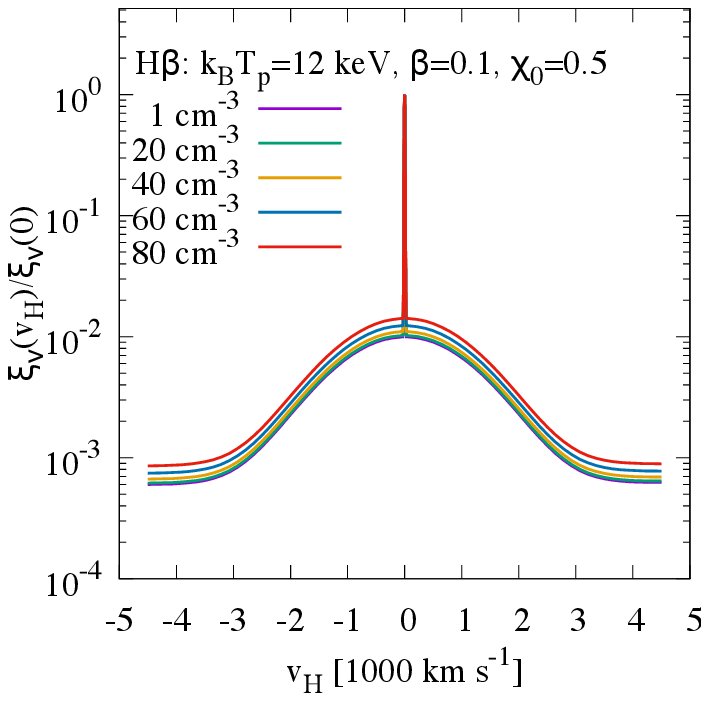}
\includegraphics[scale=0.9]{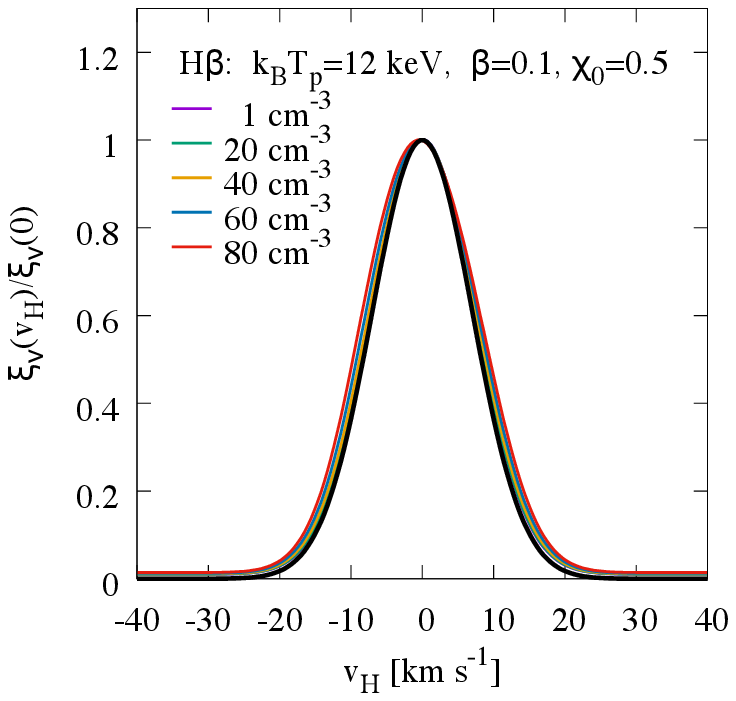}
\caption{
({\it top}):The kinetic spectra of observed H~$\beta$ for given
$k_{\rm B}T_{\rm p}=12$~keV and $\beta=0.1$.
The colours correspond to
$n_{\rm tot,0}=1~{\rm cm^{-3}}$ (black),
$n_{\rm tot,0}=20~{\rm cm^{-3}}$ (green),
$n_{\rm tot,0}=40~{\rm cm^{-3}}$ (orange),
$n_{\rm tot,0}=60~{\rm cm^{-3}}$ (blue),
and $n_{\rm tot,0}=80~{\rm cm^{-3}}$ (red).
({\it bottom}):
Close up of the narrow component.
The black line represents the Maxwellian distribution function
that we assume for the narrow hydrogen atoms.
}
\label{fig:whole-Hb}
\end{figure}

The modifications of other narrow Balmer lines are modest compared with the H~$\alpha$
because of the small absorption coefficient.
Fig.~\ref{fig:whole-Hb} shows the case of H~$\beta$.
%
%
\par
%
\begin{figure}
\centering
\includegraphics[scale=0.9]{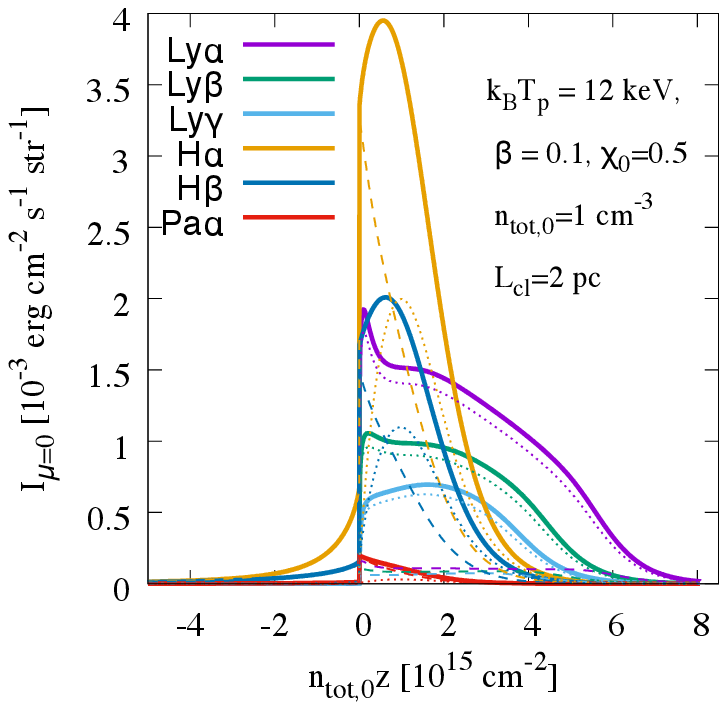}
\includegraphics[scale=0.9]{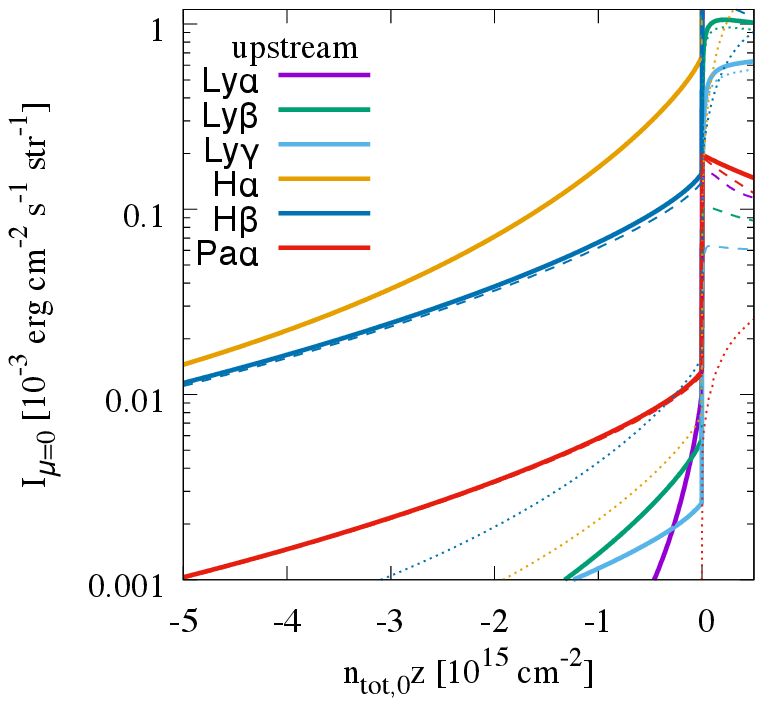}
\caption{
({\it top}):
The frequency-integrated intensities of
Ly~$\alpha$ (purple), Ly~$\beta$ (green), Ly~$\gamma$ (light blue),
H~$\alpha$ (orange), H~$\beta$ (blue) and Pa~$\alpha$ (red).
The solid line shows the total intensity.
The broken line represents the narrow component,
while the dots represent the broad component.
Here we set $k_{\rm B}T_{\rm p}=12$~keV, $\beta=0.1$, $\chi_0=0.5$
and $n_{\rm tot,0}=1~{\rm cm^{-3}}$.
({\it bottom}): Close up of the upstream region.
}
\label{fig:brightness}
\end{figure}
\begin{figure}
\centering
\includegraphics[scale=0.9]{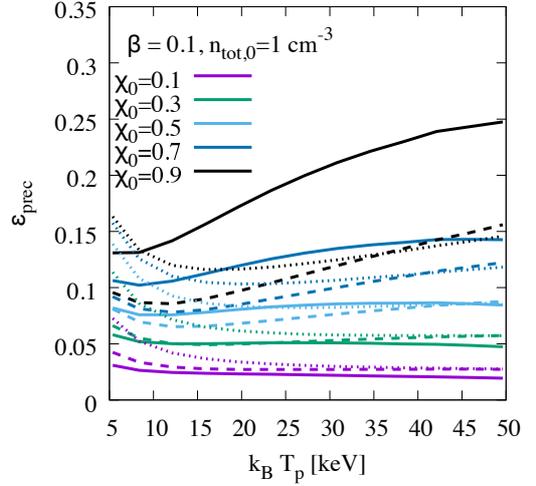}
\caption{
The intensity ratio of the photo-precursor emission to the downstream emission
for H~$\alpha$ (solid line), H~$\beta$ (broken line) and Pa~$\alpha$ (dots).
The colours correspond to
$\chi_0=0.1$ (purple),
$\chi_0=0.3$ (green),
$\chi_0=0.5$ (light blue),
$\chi_0=0.7$ (blue),
and $\chi_0=0.9$ (black).
Here we set $\beta=0.1$ and $n_{\rm tot,0}=1~{\rm cm^{-3}}$.
}
\label{fig:prec}
\end{figure}

%
Fig.~\ref{fig:brightness} shows the frequency-integrated intensity of each line,
%
\begin{eqnarray}
I_{0}(z)=\int I_{\nu,0}(z) {\rm d}\nu,
\end{eqnarray}
%
for given $k_{\rm B}T_{\rm p}=12~{\rm keV}$, $\beta=0.1$, $\chi_0=0.5$
and $n_{\rm tot,0}=1~{\rm cm^{-3}}$.
Here we also display the intensities of narrow component,
%
\begin{eqnarray}
I_{0}^{\rm N}(z)=\int_{|\nu|<|\varpi_0^{\rm N}|} I_{\nu,0}(z) {\rm d}\nu,
\end{eqnarray}
%
and broad component,
%
\begin{eqnarray}
I_{0}^{\rm B}(z)=\int_{|\nu|\ge|\varpi_0^{\rm N}|} I_{\nu,0}(z) {\rm d}\nu.
\end{eqnarray}
%
The narrow Lyman lines are mostly absorbed, while the broad Lyman lines are
transparent. The Balmer lines are also optically thin in this tenuous-gas
case ($n_{\rm tot,0}\ll30~{\rm cm^{-3}}$). The Pa~$\alpha$ is always in the
optically thin limit. There is the shock-precursor-like emission due to the
leaking of Lyman photons (see the bottom panel of Fig.~\ref{fig:brightness}).
Obviously, the length of photo-precursor emission corresponds to the mean
free path of the Lyman photons, which depends on $n_{\rm tot,0}{}^{-1}$. We
display the intensity ratio of the photo-precursor emission to the downstream
emission,
%
\begin{eqnarray}
\varepsilon_{\rm prec}
=\frac{\int_{z_{\rm out}}^{0} I_0(z) {\rm d}z}{\int_0^{z_{\rm in}} I_0(z) {\rm d}z},
\end{eqnarray}
%
for H~$\alpha$, H~$\beta$ and Pa~$\alpha$ in Fig.~\ref{fig:prec}.
%
%
The relative intensity of photo-precursor emission is increasing with
increasing ionization degree $\chi_0$ because the Lyman series lines emerging
in the downstream region tend to leak more to the upstream region. Thus, the
shock-precursor-like emission is ubiquitously observed even if there are no
leaking particles.
\par
We define the total observed intensity of the narrow component as
%
\begin{eqnarray}
\zeta_0^{\rm N}=\int_{z_{\rm out}}^{z_{\rm in}} I_0^{\rm N}(z) {\rm d}z.
\end{eqnarray}
Similarly, that of broad component is defined as
%
\begin{eqnarray}
\zeta_0^{\rm B}=\int_{z_{\rm out}}^{z_{\rm in}} I_0^{\rm B}(z) {\rm d}z.
\end{eqnarray}
%
%
\begin{figure}
\centering
\includegraphics[scale=0.9]{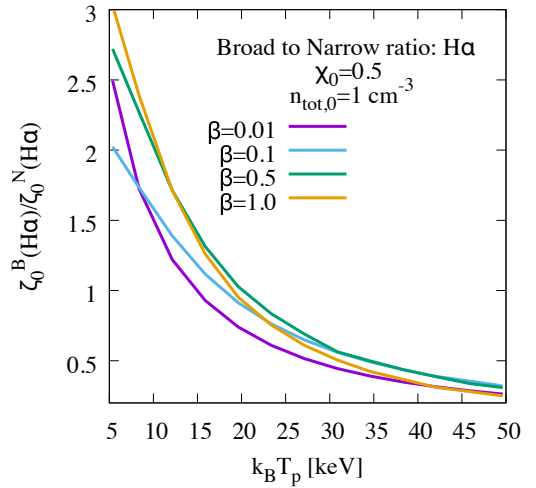}
\caption{
The intensity ratio of broad H~$\alpha$ to narrow H~$\alpha$
for given $\chi_0=0.5$ and $n_{\rm tot,0}=1~{\rm cm^{-3}}$.
The colours correspond to
$\beta=0.01$ (purple),
$\beta=0.01$ (light blue),
$\beta=0.5$ (green)
and $\beta=1$ (orange),
respectively.
}
\label{fig:Balmer-b2na-beta}
\end{figure}
%
The intensity ratio of broad H~$\alpha$ to narrow H~$\alpha$ is often relied
on to estimate $\beta$. Note that since our model simplifies the broad
hydrogen atoms, we predict only how the broad-to-narrow ratio depends on
$\chi_0$ and $\beta$. The ratio is shown in Fig.~\ref{fig:Balmer-b2na-beta}
for given $\chi_0=0.5$ and $n_{\rm tot,0}=1~{\rm cm^{-3}}$ with fixed values
$\beta_0=$(0.01, 0.1, 0.5 and 1), while Fig.~\ref{fig:Balmer-b2na-xion} shows
the ratio for given $\beta=0.1$ and $n_{\rm tot,0}=1~{\rm cm^{-3}}$ with
fixed values of $\chi_0=$(0.1,0.5 and 0.9).
%
\begin{figure}
\centering
\includegraphics[scale=0.9]{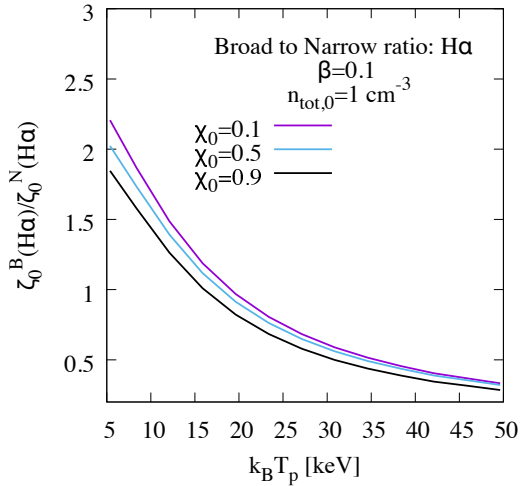}
\caption{
The intensity ratio of broad H~$\alpha$ to narrow H~$\alpha$
for given $\beta=0.1$ and $n_{\rm tot,0}=1~{\rm cm^{-3}}$.
The colours correspond to
$\chi_0=0.1$ (purple),
$\chi_0=0.5$ (light blue),
$\chi_0=0.7$ (blue),
and $\chi_0=0.9$ (black).
}
\label{fig:Balmer-b2na-xion}
\end{figure}
%
The broad-to-narrow ratio of H~$\alpha$ also depends on $\chi_0$
due to the conversion of narrow Ly~$\beta$ to narrow H~$\alpha$
but the dependence on $\chi_0$
is modest compared with the case of $\beta$ at lower $T_{\rm p}$.
\par
Because of the mildly high opacity of H~$\alpha$ for $n_{\rm tot,0}\ga10~{\rm cm^{-3}}$,
the peak of narrow H~$\alpha$ is reduced in comparison of the case of $n_{\rm tot,0}=1~{\rm cm^{-3}}$.
Therefore, the intensity ratio of broad H~$\alpha$ to narrow H~$\alpha$ depends on the ambient density.
%
\begin{figure}
\centering
\includegraphics[scale=0.9]{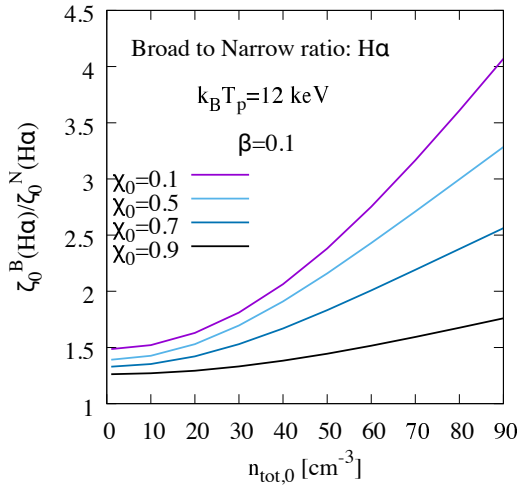}
\caption{
The intensity ratio of broad H~$\alpha$ to narrow H~$\alpha$
for given $k_{\rm B}T_{\rm p}=12~{\rm keV}$ and $\beta=0.5$.
The colours correspond to
$\chi_0=0.1$ (purple),
$\chi_0=0.5$ (light blue),
$\chi_0=0.7$ (blue),
and $\chi_0=0.9$ (black).
}
\label{fig:Balmer-b2na-density}
\end{figure}
%
Fig.~\ref{fig:Balmer-b2na-density} shows the broad-to-narrow ratio of H~$\alpha$
as function of $n_{\rm tot,0}$ for given $k_{\rm B}T_{\rm p}=12~{\rm keV}$ and $\beta=0.1$
with fixed values $\chi_0=$(0.1, 0.5, 0.7 and 0.9).
\par
%
\begin{figure}
\centering
\includegraphics[scale=0.9]{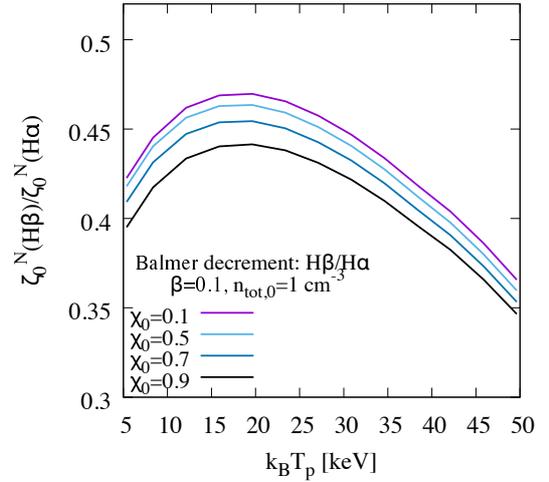}
\caption{
The intensity ratio of narrow H~$\beta$ to H~$\alpha$
for given $\beta=0.5$ and $n_{\rm tot,0}=1~{\rm cm^{-3}}$.
The colours correspond to
$\chi_0=0.1$ (purple),
$\chi_0=0.5$ (light blue),
$\chi_0=0.7$ (blue),
and $\chi_0=0.9$ (black).
Note that the ratio of photon counts shown in~\citet{shimoda18}
is given by $(\nu_{\rm 3p,2s}'/\nu_{\rm 4p,2s}')\zeta_0^{\rm N}({\rm H\beta})/
\zeta_0^{\rm N}({\rm H\alpha})
\simeq0.74\zeta_0^{\rm N}({\rm H\beta})/
\zeta_0^{\rm N}({\rm H\alpha})$.
}
\label{fig:Balmer-decrement-xion}
\end{figure}
%
Fig.~\ref{fig:Balmer-decrement-xion} shows the ratio of the total narrow
intensity of H~$\beta$ to that of H~$\alpha$, $\zeta_0^{\rm N}({\rm
H\beta})/\zeta_0^{\rm N}({\rm H\alpha})$, i.e. the Balmer decrement.
Note that the ratio of photon counts shown in~\citet{shimoda18} is
given by $(\nu_{\rm 3p,2s}'/\nu_{\rm 4p,2s}')\zeta_0^{\rm N}({\rm H\beta})/
\zeta_0^{\rm N}({\rm H\alpha}) \simeq0.74\zeta_0^{\rm N}({\rm H\beta})/
\zeta_0^{\rm N}({\rm H\alpha})$\footnote{ The Balmer decrement shown
in~\citet{shimoda18} was implicitly defined by $I_{\nu}/h\nu$ to compare with
the observation by~\citet{sparks15}. }.
The ratio depends on how many Ly~$\beta$ and Ly~$\gamma$ photons are converted to
H~$\alpha$ or H~$\beta$ photons. Since the escape fraction of Ly~$\beta$,
$\varepsilon_{\rm esc}({\rm Ly\beta})$, depends strongly on the ionization
degree $\chi_0$ compared to the fraction of $\rm Ly\gamma$ (see
Fig.~\ref{fig:trap}), in other words $\left|\frac{\upartial
\zeta_0^{\rm N}({\rm H\alpha})}{\upartial \chi_0}\right|>\left|\frac{\upartial
\zeta_0^{\rm N}({\rm H\beta})}{\upartial \chi_0}\right|$,
a larger $\chi_0$ results in a less conversion of
Ly~$\beta$ to H~$\alpha$ than the case of Ly~$\gamma$ to H~$\beta$
conversion. Thus, the observed ratio $\zeta_0^{\rm N}({\rm
H~\beta})/\zeta_0^{\rm N}({\rm H~\alpha})$ would decrease with increasing
$\chi_0$. On the other hand, the ratio also depends
on the interval of spatial integration of Eq.~\eqref{eq:xi}. For example,
when we take the interval from $z=0$ to $z=z_{\rm in}$ (i.e. observe the downstream
region), the ratio increases with increasing $\chi_0$, with values around $0.4\mathchar`-0.5$.
%
\begin{figure}
\centering
\includegraphics[scale=0.9]{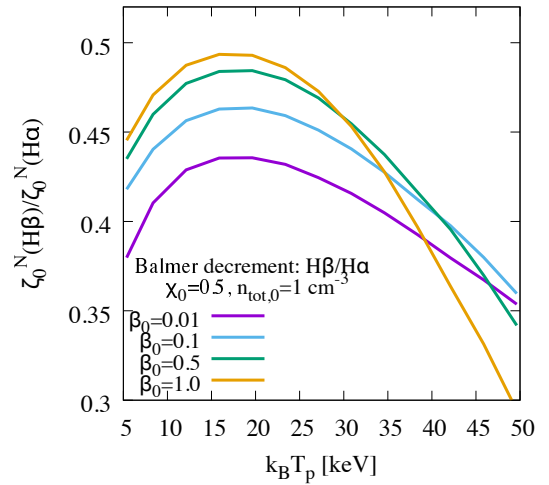}
\caption{
The intensity ratio of narrow H~$\beta$ to H~$\alpha$
for given $\chi_0=0.5$ and $n_{\rm tot,0}=1~{\rm cm^{-3}}$.
The colours correspond to
$\beta=0.01$ (purple),
$\beta=0.1$ (light blue),
$\beta=0.5$ (green)
and $\beta=1$ (orange),
respectively.
Note that the ratio of photon counts shown in~\citet{shimoda18}
is given by $(\nu_{\rm 3p,2s}'/\nu_{\rm 4p,2s}')\zeta_0^{\rm N}({\rm H\beta})/
\zeta_0^{\rm N}({\rm H\alpha})
\simeq0.74\zeta_0^{\rm N}({\rm H\beta})/
\zeta_0^{\rm N}({\rm H\alpha})$.
}
\label{fig:Balmer-decrement-beta}
\end{figure}
%
Fig.~\ref{fig:Balmer-decrement-beta} shows the Balmer decrement for given
$\chi_0=0.5$ and $n_{\rm tot,0}=1~{\rm cm^{-3}}$ with fixed values of
$\beta=$(0.01, 0.1, 0.5 and 1). The ratio varies with various $\chi_0$ and
$\beta$ with $\sim10$~per cent. Note that the values of
ratio around $0.4\mathchar`-0.5$ seem to be consistent with observed value in
SN~1006, $\sim0.5\pm0.1$~\citep{raymond17}. On the other hand, the ratios
$\la0.3$ are observed by~\citet{ghavamian01,ghavamian02} for {\it
Tycho's}~SNR and SN~1006 (at different position from~\citet{raymond17}). The
small value of ratio could be explained if the SNR shock suffers extreme
energy losses~\citep[][see the results of $\eta>0.5$]{shimoda18}.
\par
The difference of absorption coefficients between H~$\alpha$ and H~$\beta$
leads to a variation of the Balmer decrement
$\zeta_0^{\rm N}({\rm H\beta})/\zeta_0^{\rm N}({\rm H\alpha})$
with the ambient density $n_{\rm tot,0}$.
%
\begin{figure}
\centering
\includegraphics[scale=0.9]{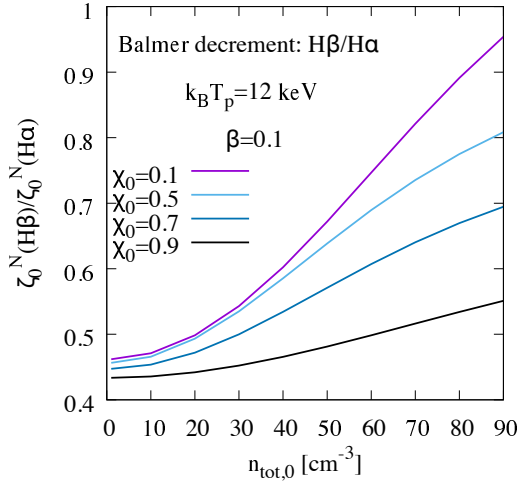}
\caption{
The intensity ratio of narrow H~$\beta$ to H~$\alpha$
for given $k_{\rm B}T_{\rm p}=12~{\rm keV}$ and $\beta=0.5$.
The colours correspond to
$\chi_0=0.1$ (purple),
$\chi_0=0.5$ (light blue),
$\chi_0=0.7$ (blue),
and $\chi_0=0.9$ (black).
}
\label{fig:Balmer-decrement-density}
\end{figure}
%
Fig.~\ref{fig:Balmer-decrement-density} shows the Balmer decrement as
function of $n_{\rm tot,0}$ for the given $k_{\rm B}T_{\rm p}=12~{\rm keV}$
and $\beta=0.1$ with fixed values $\chi_0=$(0.1, 0.5, 0.7 and 0.9). Note that
both variations of $\zeta_0^{\rm N}({\rm H\beta})/\zeta_0^{\rm N}({\rm
H\alpha})$~(see Fig.~\ref{fig:Balmer-b2na-density}) and $\zeta_0^{\rm B}({\rm
H\alpha})/\zeta_0^{\rm N}({\rm H\alpha})$ on $n_{\rm tot,0}$ result from the
scattering of narrow H~$\alpha$. Therefore, they resemble each other.

\section{Summary and Discussion}
\label{sec:discussion} We have studied the radiative transfer of hydrogen
lines taking into account the atomic population. The absorption of Lyman
photons results in the emission of Balmer photons and yields 2s-state
hydrogen atoms. We have shown that a fraction of the Lyman photons escape
from the shock toward the far downstream region and therefore the
Lyman-Balmer conversion is incomplete, which is consistent with the results
of~\citet{ghavamian01}. As a result, the observed intensity ratio of narrow
H~$\beta$ to narrow H~$\alpha$ (the Balmer decrement) and the observed
broad-to-narrow ratio of H~$\alpha$ vary $\sim10$ per cent depending on
the ionization degree of upstream medium. Note that the broad-to-narrow ratio
of H~$\alpha$ is often relied upon to derive the ion-electron temperature
equilibrium in the downstream region rather than the ionization degree.
The intensity ratios depend also on the ambient density because of the
scattering of H~$\alpha$ by the 2s-state hydrogen atoms if the shock propagates
into the medium with density of $\ga10~{\rm cm^{-3}}$. In the case of optically
thick H~$\alpha$, the line shape of narrow H~$\alpha$ becomes asymmetric as
a consequence of the differences in atomic energy levels. The degree of
asymmetry increases with increasing the optical depth. Since the narrow
H~$\alpha$ can be scattered, the observed Balmer decrement and broad-to-narrow
depend also on the ambient density.
\par
We have supposed that the H~$_{\rm I}$ cloud in the path of the shock is
clearly separated into regions of dense clumps and a diffuse, tenuous
component and that the shock-fronts propagating into each part are isolated.
Moreover, we have fixed the temperature for H~$_{\rm I}$ gas. However, the
temperature depends on the density in reality~\citep[e.g.][]{field65}. The
clumpy gas (`cold neutral medium') is usually supposed to have the
temperature of $\sim50\mathchar`-100$~K with the density of
$\sim10\mathchar`-100~{\rm cm^{-3}}$, while the diffuse gas (`warm neutral
medium') is supposed to have the temperature of $\sim6000\mathchar`-10000$~K
with the density of $\sim0.2\mathchar`-0.5~{\rm cm^{-3}}$~\citep[e.g.][and
references therein]{ferriere01}. On the other hand, motivated by
investigations of the molecular cloud formation via accretion flows of
H~$_{\rm I}$ clouds (i.e. star formation), recent numerical simulations show
that during the formation, the density and temperature of H~$_{\rm I}$ cloud
range over two to three orders of magnitude~\citep[e.g.][see also Figures~5
and 6 of \citealt{fukui18}] {inoue12,inoue16}. The probability distribution
function in the density-temperature plane presented in~\citet{fukui18} shows
the existence of H~$_{\rm I}$ clouds with density of
$\sim10\mathchar`-100~{\rm cm^{-3}}$ and temperature of
$\sim10^3\mathchar`-10^4~{\rm K}$. The size of such a cloud is $\sim1~{\rm
pc}$, while the size of denser clouds ($\ga100~{\rm cm^3}$ and $\sim100~{\rm
K}$) is $\sim0.1$~pc. The separation of these clouds is typically $\sim1~{\rm
pc}$, but sometimes $\sim0.1~{\rm pc}$. Thus, for the case of actual SNR
shocks, the densest clump, tenuous gas and intermediately dense cloud can be
co-existing on our line of sight. Note that such H~$_{\rm I}$ clouds
associated with the molecular cloud formation may be a minor component of the
ISM, but it may be possible that the SNR shock propagates in such H~$_{\rm
I}$ regions because of the very long dynamical time of ISM, $\sim(10~{\rm
pc}/10~{\rm km~s^{-1}})\sim10^6~{\rm yr}$.

\par
We have not considered any other radiation sources, especially the radiation
from supernova ejecta. Moreover, we neglect the leaking of broad hydrogen
atoms and the existence of cosmic-rays. The cosmic-rays
or neutral leakage precursors are accelerated relative to the pre-shock gas,
giving a velocity offset as an extra parameter in the radiative transfer
calculation~\citep{boulares88}. Note that the broad hydrogen atoms leaking
to the upstream region may lead to the modification of shock structure as
well as that arising in a cosmic-ray modified
shock~\citep[][]{blasi12,ohira12,ohira13,ohira16,ohira16b}. Furthermore, to
quantify the production of non-thermal particles, polarization measurements
are additionally required~\citep{shimoda18}. We will address these issues
(realistic ISM, radiation from supernova ejecta, leaking particles and
polarization) in forthcoming papers. In principle, the 2s$_{1/2}$ population
may also be quenched by mixing with the 2p$_{1/2}$ level
in a motional electric field~\citep[e.g.][]{drake88}. In an electric field of
$E=10^{-6}$~StatVolt~cm$^{-1}$, arising from motion with a velocity
3000~km~s$^{-1}$ through a magnetic field of $10^{-4}$~G, the mixing
amplitude evaluates to $2\times 10^{-6}\left(E/10^{-6}\right)$, leading to a
negligible increase in the decay rate of $\left(2\times
10^{-6}\right)^2\times 6.25\times 10^8 = 0.0025$ s$^{-1}$. A factor of 60
increase in electric field is required to make this new contribution
comparable with the two-photon decay rate of $\sim 8$ s$^{-1}$, which for now
seems to be out of reach for plausible cosmic ray generated magnetic fields
and SNR shock velocities \citep[e.g.][]{vink03}.
\par
Although our model is still too simple to compare line profiles
quantitatively with actual observations, it can predict {\it different} line
profiles between H~$\alpha$ and H~$\beta$. This qualitative prediction is
based on the atomic physics (i.e. oscillator strength $f_{j,k}$) rather than
the property of BDSs. In particular, if the anomalous width of H~$\alpha$
reflects the intrinsic velocity dispersion of hydrogen atoms, both of
H~$\alpha$ and H~$\beta$ should have the same width. On the other hand, in
the case of scattering, the width should depend on the direction of our line
of sight (i.e. optical depth). For $\mu\simeq0$ ($\mu\simeq1$), the optical
depth tends to be maximum (minimum). Interestingly, the width of narrow
H~$\alpha$ observed in SNR~0509~67.5 depends on the line of
sight~\citep{smith94}. The width observed at the west rim of SNR
($\mu\simeq0$) is $\simeq30\pm2~{\rm km~s^{-1}}$, while at the centre of SNR
($\mu\simeq1$) is $\simeq25\pm2~{\rm km~s^{-1}}$. It is qualitatively
consistent with the difference of optical depth. Note that \citet{smith94}
suggested the existence of intermediate component for the west rim. This
scenario can be tested by future observations of H~$\beta$ with
high-resolution spectroscopy.
\par
\citet{long92} reported the two-photon continuum emission
from SNR~Cygnus~Loop at which $k_{\rm B}T_{\rm
p}\sim0.1$~keV~\citep{medina14}. Unfortunately, since there are no
cross-section data on proton impact excitation to $n_k=4$ at a relative
velocity less than $1000~{\rm km~s^{-1}}$, our present calculation is limited
at $k_{\rm B}T_{\rm p}\ga5$~keV. According to Figs.~\ref{fig:whole-Ha} and
\ref{fig:whole-Hb}, the two-photon continuum can be a diagnostic of BDSs. To
do this, we should treat the broad hydrogen atoms
self-consistently~\citep[e.g., see][]{raymond08,blasi12,ohira12} and consider
any other radiation sources that yield continuum components. We will extend
our model to provide a diagnostic from the two-photon continuum in future
work. 

\section*{Acknowledgements}


We thank Dr. Makito Abe for valuable comments that helped us complete this
work. We also thank the referee, John Raymond, for his
comments further improve the paper.
This work is partially supported by JSPS KAKENHI grant no. JP18H01245.
JML was supported by the Guest Investigator Grant HST-GO-13435.001 from the
Space Telescope Science Institute and by the NASA Astrophysics Theory Program
(80HQTR18T0065), as well by Basic Research Funds of the CNR.




\bibliographystyle{mnras}
\bibliography{mnras_LineTransfer}




\appendix





\bsp	
\label{lastpage}
\end{document}